\documentclass[article,onecolumn,superscriptaddress,nofootinbib]{revtex4}
\usepackage[dvipdfmx]{graphicx}
\usepackage{amsfonts,amsmath,amssymb,epsf}
\usepackage{amsmath,braket}
\usepackage{graphicx,color}
\usepackage{bm}
\usepackage{graphicx}
\usepackage{amstext}
\usepackage{dsfont}
\usepackage{float}
\usepackage[usenames,dvipsnames]{xcolor}
\usepackage[colorlinks=true,citecolor=Cerulean,linkcolor=RubineRed,urlcolor=Cerulean]{hyperref}

\newcommand{\be}{\begin{equation}}
\newcommand{\ba}{\begin{eqnarray}}
\newcommand{\ea}{\end{eqnarray}}
\newcommand{\ee}{\end{equation}}

\newcommand{\s}{\sqrt}

\newcommand{\ddd}{\cdot\cdot\cdot}
\newcommand{\no}{\nonumber \\}
\newcommand{\la}{\langle}
\newcommand{\lb}{\rangle}
\newcommand{\bea}{\begin{eqnarray}}
\newcommand{\eea}{\end{eqnarray}}
\newcommand{\bes}{\begin{equation*}}
\newcommand{\beas}{\begin{eqnarray*}}
\newcommand{\eeas}{\end{eqnarray*}}
\newcommand{\bas}{\begin{array*}}
\newcommand{\eas}{\end{array*}}
\newcommand{\ees}{\end{equation*}}

\def\ee{{\rm e}}

\def\la{\langle}
\def\ra{\rangle}

\def\tr{{\rm tr}}

\newcommand{\beq}{\begin{equation}}
\newcommand{\eeq}{\end{equation}}
\newcommand{\beqa}{\begin{eqnarray}}
\newcommand{\eeqa}{\end{eqnarray}}

\numberwithin{equation}{section}

\begin{document}

\begin{flushright}
YITP-19-106
\\
IPMU 19-0167
\\
\end{flushright}

\title{Decoherence  in Conformal Field Theory}

\author{Adolfo del Campo}
\affiliation{Donostia International Physics Center,  E-20018 San Sebasti\'an, Spain}
\affiliation{IKERBASQUE, Basque Foundation for Science, E-48013 Bilbao, Spain}
\affiliation{Department of Physics, University of Massachusetts, Boston, MA 02125, USA}

\author{Tadashi Takayanagi}
\affiliation{Center for Gravitational Physics, Yukawa Institute for Theoretical Physics, Kyoto University, Kyoto 606-8502, Japan}
\affiliation{Kavli Institute for the Physics and Mathematics of the Universe (WPI), University of Tokyo, Kashiwa, Chiba 277-8582, Japan}

\begin{abstract}
Noise sources are ubiquitous in Nature and give rise to a description of quantum systems in terms of stochastic Hamiltonians. Decoherence dominates the noise-averaged dynamics and leads to dephasing and the decay of coherences in the eigenbasis of the fluctuating operator. For energy-diffusion processes stemming from fluctuations of the system Hamiltonian the characteristic decoherence time is shown to be proportional to the heat capacity. We analyze the decoherence dynamics of entangled CFTs and characterize the dynamics of the purity, and logarithmic negativity, that are shown to decay monotonically as a function of time. The converse is true for the quantum Renyi entropies. From the short-time asymptotics of the purity, the decoherence rate is  identified and shown to be proportional to the central charge. The fixed point  characterizing long times of evolution depends on the presence degeneracies in the energy spectrum. We show how information loss associated with decoherence can be attributed to its leakage to an auxiliary environment and  discuss how gravity duals of decoherence dynamics in holographic CFTs looks like in AdS/CFT. We find that the inner horizon region of eternal AdS black hole is highly squeezed due to decoherence. 
\end{abstract}

\maketitle

\newpage

\tableofcontents

\newpage

\section{Introduction}

Decoherence leads to the emergence of classical behavior in a quantum word. Decoherence is indeed an ubiquitous phenomenon, stemming  from the entanglement of a quantum system with its  surrounding environment \cite{Zurek03}.
Its description is natural in terms of a composite large system including both system and environment.
Due to the difficulty to describe the dynamics in such setting, (master) equations of motion governing the reduced density matrix of the system of interest are often employed. 

Prominent aspects  of decoherence involve the inclusion of noise in the degrees of freedom of the system.
Noise in the energy spectrum can be described by  fluctuations in time in the system Hamiltonian, that lead to energy-diffusion processes. As it turns out, such kind of noise is of relevance in a wide variety of contexts. 
Within standard quantum mechanics, if the evolution of the system is described using a clock, 
errors in the clock time keeping give rise to fluctuations in the system Hamiltonian \cite{Egusquiza99,Gambini07}.
A  different setting in which the system Hamiltonian is also a fluctuating operator is associated with wavefunction collapse models \cite{Gisin84,Percival94,Adler03,Bassi03,Bassi13}. The latter postulate  
that the fundamental equation of motion of an isolated quantum system is not the (deterministic) Schr\"odinger equation but its stochastic generalization with a single fluctuating term $V=H$. This feature is shared by other modifications of quantum mechanics, e.g, entertaining the possibility that the phase of the quantum state fluctuates in a short-time scale \cite{Milburn91}.
Yet, other scenarios characterized by fluctuating Hamiltonians concern quantum simulation protocols in which noise is harnessed as a resource for the emulation of open quantum systems
 \cite{Chenu17} and random measurement Hamiltonians \cite{Korbicz17}.

With the emergence of classical behavior, quantum correlations such as entanglement are suppressed.
Entanglement plays a crucial role  in quantum matter and  is essential in the characterization of quantum field theories.
In particular, it has played a key role in holography, that relates a conformal field theory (CFT) to a quantum theory of gravity in anti-de Sitter space (AdS) \cite{Maldacena99}. In this context, entanglement is key to the emergence of gravity in CFTs and several measures of entanglement admit a geometric interpretation in AdS. Remarkably, the entanglement entropy can be computed as a minimal area in AdS \cite{RT06,RT06b,Hubeny07}. Analogous relations hold for a variety of measures of quantum correlations (so called entanglement of purification) \cite{Umemoto18,Nguyen18}.
It is thus imperative to acquire an understanding of how decoherence affects field theories in the context of AdS/CFT. 

Motivated by these we will study the decoherence in CFTs induced by fluctuating Hamiltonian $V=H$.
This energy fluctuations can be attributed to noise in the clock time in CFTs or equally to noise in the time component of the metric of CFTs. Therefore this is the most fundamental class of noises for CFTs and their gravity duals.

\section{Noise-induced Decoherence}
A noisy quantum system  is described by a fluctuating Hamiltonian
\beqa
H_T=H_{0}+\hbar \sum_{\mu }\sqrt{\gamma _{\mu }}\xi _{t}^{\mu }V_{\mu },
\eeqa
where $\xi _{t}^{\mu }$ represent independent Gaussian random processes with zero-mean $\langle \xi _{t}^{\mu }\rangle=0$ and white noise correlations 
$\langle \xi _{t}^{\mu }\xi _{t'}^{\nu }\rangle=\delta_{\mu\nu}\delta(t-t')$.
For a given set of realizations of these processes $\{\xi _{t}^{\mu }\}$, the dynamics is governed by the stochastic master equation 
\begin{eqnarray}
\mathrm{d}|\Psi _{t}^{\xi }\rangle &=&-\frac{i}{\hbar }H_{0}\mathrm{d}t|\Psi
_{t}^{\xi }\rangle -i\sum_{\mu }\sqrt{\gamma _{\mu }}V_{\mu }\mathrm{d}%
W_{t}^{\mu }|\Psi _{t}^{\xi }\rangle  -\frac{1}{2}\sum_{\mu }\gamma _{\mu
}V_{\mu }^{2}\mathrm{d}t|\Psi _{t}^{\xi }\rangle ,
\end{eqnarray}
where $\mathrm{d}W_{t}^{\mu }$, defined from $\xi _{t}^{\mu }:=\mathrm{d}%
W_{t}^{\mu }/\mathrm{d}t$ is an It\^{o} stochastic differential. As such, it obeys the relations
$\mathrm{d}W_{t}^{\mu }\mathrm{d}%
W_{t}^{\nu }=\delta _{\mu \nu }\mathrm{d}t$ and $\mathrm{d}W_{t}^{\mu }%
\mathrm{d}t=\mathrm{d}t^{2}=0$.
The noise-averaged dynamics is governed by the master equation
\begin{equation}
\dot{\rho}(t)=-\frac{i}{\hbar }[H_{0},\rho(t)]-\frac{1}{2}\sum_{\mu
}\gamma _{\mu }\left[ V_{\mu },\left[ V_{\mu },\rho(t)\right] \right],
\label{ME}
\end{equation}
where the density matrix represents the average over different realizations of the noise
 $\rho(t)=\langle \rho(t)^{\xi }\rangle _{\xi
} $.
The evolution (\ref{ME}) is Markovian (memory-less) and  of  Lindblad form with Hermitian Lindblad operators $V_{\mu }=V_{\mu }^\dag$ \cite{Lindblad76}. 
Relevant characteristic times in Markovian open quantum dynamics can often be identified from the early-time behavior \cite{Chenu17,Beau17}.
In \cite{Xu19}, it was shown that the decoherence rate $D$ can be extracted from the short-time asymptotics of the purity $P(t)=\tr[\rho(t)^2]$, i.e.,
\beqa
\label{sapurity}
P(t)=P(0)(1-Dt)+\mathcal{O}(t^2),
\eeqa
which is consistent with an exponential behavior to order $\mathcal{O}(t)$.
The decoherence rate reads
\beqa
D=\frac{2}{P(0)}\sum_{\mu}\gamma _{\mu } {\rm } \widetilde{\mathrm{var}}_{\rho(0)}(V_\mu),
\eeqa
where the modified variance reads $\widetilde{\mathrm{var}}_{\rho(0)}(X):=\left\langle\rho(0)X^{2}\right\rangle _{\rho(0)}-\left\langle X\rho(0)X\right\rangle
_{\rho(0)}$, with $\left\langle \cdot \right\rangle _{\rho(0)}:=$tr$%
\left( \rho(0)\cdot \right)$.

As a function of the system size, $D$ exhibits at most  a polynomial dependence on the system size whenever the fluctuating operator is $k$-local.
By contrast, whenever the fluctuating operator is fully non-local  (e.g., a random matrix operator), the decoherence rate becomes proportional to the Hilbert space dimension, 
and this exhibits an exponential dependence on systems of interacting qubits \cite{Xu19}.

\section{Decoherence of Entangled Quantum States}

Consider a canonical Gibbs state associated with a system at thermal equilibrium at inverse temperature $\beta =(k_{B}T)^{-1}$, where $k_B$ is the Boltzmann constant.
The state of the system is described by a density matrix $\rho_\beta=e^{-\beta H}/Z(\beta )$, where the partition function equals $Z(\beta ):=%
\mathrm{tr}(e^{-\beta H})$. Let the spectral decomposition of the system Hamiltonian be given by $H=\sum_kE_k|k\rangle\langle k|$, 
in terms of the energy eigenvalues $E_k$ and the corresponding eigenstates $|k\rangle$. We use the subindex $k$ as a global quantum number that accounts for possible degeneracies in the spectrum.

A purification of this state can be achieved by doubling the Hilbert space
and considering two identical copies of the system. The resulting
 thermofield double (TFD) state \cite{Umezawa82}%
\begin{equation}
\left\vert {\rm TFD}\right\rangle :=\frac{1}{\sqrt{Z(\beta )}}\sum_{k}e^{-%
\frac{\beta E_{k}}{2}}\left\vert k\right\rangle \left\vert k\right\rangle .
\label{TFD}
\end{equation}%
is an entangled state of the two copies.   The reduced density matrix obtained by tracing over any of the two copies of the system equals the canonical thermal state $\rho_\beta$.

TFD states are commonly used in
finite-temperature field theory and have been widely studied in the context
of holography, e.g., in connection to the entanglement between black holes~\cite{Maldacena13}, the butterfly effect~ \cite{Shenker14}, and quantum
source-channel codes~\cite{Pastawski17}.

 Assume now that each subsystem is perturbed by a
single Gaussian real white noise $\xi _{t}$ \cite{Nagasawa00}, so that  the total Hamiltonian is given by 
\begin{equation}
{H}_T=H\otimes \mathds{1}+%
\mathds{1}\otimes H+\hbar \sqrt{\gamma }(\xi _{t}^{L}H\otimes \mathds{1}+%
\mathds{1}\otimes \xi _{t}^{R}H), 
\end{equation}%
where we assume independent noises acting on each copy, $\xi
_{t}^{L}\neq \xi _{t}^{R}$, with identical amplitudes $\gamma _{L}=\gamma
_{R}=\gamma $. The dynamics of the system is governed by the Schr\"{o}dinger
equation $i\hbar \partial _{t}|\Psi _{t}^{\xi }\rangle =H_T|\Psi
_{t}^{\xi }\rangle $, denoting $\xi :=\{\xi _{t}^{L},\xi _{t}^{R}\} $ for
simplicity.

As shown in \cite{Xu19}, the noise-averaged density
matrix $\rho(t)=\langle |\Psi _{t}^{\xi }\rangle \langle \Psi _{t}^{\xi
}|\rangle _{\xi }$ is governed by the master equation~\eqref{ME}, that is of Lindblad form \cite{Lindblad76} with Hermitian
Lindblad operators 
\begin{eqnarray}
\label{linbladops}
V_{L}=H\otimes \mathds{1}, \quad V_{R}=\mathds{1}\otimes H.
\end{eqnarray}
The decoherence rate is set by the sum of the thermal energy fluctuations of each copy, 
\begin{eqnarray}
D&=&4\gamma \mathrm{var}_{\rho _{\beta }}(H)=4\gamma \frac{\mathrm{d}%
^{2}}{\mathrm{d}\beta ^{2}}\ln \left[ Z(\beta )\right] \nonumber
\\&=&4\frac{\gamma}{k_B\beta^2} C,  
\end{eqnarray}
where  $C$ is  the heat capacity of a single copy of the system at thermal equilibrium with inverse temperature $\beta$.

\section{Thermofield Double State  under Decoherence}
The exact time evolution  under noise-induced decoherence was recently discussed in chaotic quantum systems described by fluctuating random-matrix Hamiltonians \cite{Xu19}. We emphasize here the system-independent features.

Consider a thermofield double state as  the initial state 
\begin{eqnarray}
\rho(0) &=&|{\rm TFD}\rangle \langle {\rm TFD}|  \notag \\
&=&\frac{1}{Z(\beta )}\sum_{k,\ell }e^{-\frac{\beta }{2}(E_{k}+E_{\ell
})}|k\rangle |k\rangle \langle \ell |\langle \ell |,
\end{eqnarray}%
for a given single-copy Hamiltonian $H$, with eigenvectors $|k\ra$ and  eigenvalues $E_{k}$,  satisfying $H|k\ra=E_k|k\ra$. 
The time-evolution of the density matrix can be obtained in a closed form as its matrix elements fulfill
\begin{equation}
\dot{\rho}_{kk,\ell \ell }=\frac{2}{i\hbar }(E_{k}-E_{\ell }){\rho }%
_{kk,\ell \ell }-\gamma (E_{k}-E_{\ell })^{2}{\rho }_{kk,\ell \ell }.
\end{equation}%
Upon integration, the time-evolving state is found to be given by
\begin{eqnarray}  \label{eq:SMevolution}
\rho(t) &=&\sum_{k,\ell }{\rho }_{kk,\ell \ell }(t=0)e^{-i\frac{2t}{\hbar }%
(E_{k}-E_{\ell })-\gamma t (E_{k}-E_{\ell })^{2}}|k\rangle |k\rangle \langle
\ell |\langle \ell |  \notag \\
&=&\frac{1}{Z(\beta )}\sum_{k,\ell }e^{-\frac{\beta }{2}(E_{k}+E_{\ell
})}e^{-i\frac{2t}{\hbar }(E_{k}-E_{\ell })-\gamma t(E_{k}-E_{\ell
})^{2}}|k\rangle |k\rangle \langle \ell |\langle \ell |.
\end{eqnarray}
Making use of the Hubbard-Stratanovich transformation 
\begin{equation}
e^{-t\gamma (E_{k}-E_{\ell })^{2}}=\sqrt{\frac{1}{4\pi \gamma t}}%
\int_{-\infty }^{\infty }e^{-\frac{y^{2}}{4\gamma t}}e^{\mp iy(E_{k}-E_{\ell })}%
\mathrm{d}y,
\end{equation}%
we can write the time-dependent density matrix in the form
\begin{eqnarray}
\label{rhotHS}
\rho(t) &=&\frac{1}{Z(\beta )}\left(\frac{1}{4\pi\gamma t}\right)^{\frac{1}{2}}\int \mathrm{d}ye^{-\frac{y^{2}
}{4\gamma t}}
\sum_{k,\ell }e^{-\left(\frac{\beta }{2}+i\frac{2t}{\hbar }+iy\right)E_{k}}e^{-\left(\frac{\beta }{2}-i\frac{2t}{\hbar }-iy\right)E_{\ell}}|k\rangle |k\rangle \langle \ell |\langle \ell |.
\end{eqnarray}

We note that during this time-evolution, diagonal elements of the density matrix in the energy eigenbasis are unaffected and thus energy moments
$\langle (H\otimes \mathds{1}+%
\mathds{1}\otimes H)^n\rangle$ are preserved as a a function of time and set by ($2^n$ times) the corresponding thermal value.
Similarly, the reduced density matrix obtained by tracing over the degrees of freedom of any of the copies is unaffected by decoherence, as the canonical thermal state commutes with the fluctuating operator, i.e., the system Hamiltonian.

In the absence of degeneracies in the spectrum, the fixed-point of the evolution (\ref{rhotHS}) is given by
\begin{equation}
\label{rholongt}
\rho(\infty)=\frac{1}{Z(\beta )}\sum_{k}e^{-\beta E_{k}}|k\rangle
|k\rangle \langle k|\langle k|,
\end{equation}%
that is a diagonal state in the energy eigenbasis containing classical correlations between the two copies. Thus, as $t\rightarrow \infty $ the off-diagonal
elements (so-called coherences) of the density matrix decay to zero, showing
that entanglement is lost under the decoherent dynamics.

In the process, the state becomes mixed. The degree of mixedness  is well quantified by the purity $P(t)=\tr[\rho(t)^2]$, satisfying
 $\frac{1}{d} \leq P(t)\leq 1$, where $d$ is the dimension of the Hilbert space.
The purity of the time-dependent density matrix at all times of evolution reads 
\begin{equation}
\label{purityt}
P(t)=\frac{1}{Z(\beta )^{2}}\sum_{k,\ell }e^{-\beta (E_{k}+E_{\ell
})-2\gamma t(E_{k}-E_{\ell })^{2}}.
\end{equation}%
Clearly, this is a function that decays monotonically as a function of time. This feature is a direct consequence of the unital character of the energy-dephasing dynamics. Indeed, the master equation (\ref{ME}) is of Lindblad form \cite{Lindblad76} with Hermitian Lindblad operators, a sufficient condition for the dynamics to be unital  \cite{Lidar06}.
From the expression (\ref{purityt}), it is apparent that
\beqa
\frac{Z(2\beta )%
}{Z(\beta )^{2}}
\leq P(t)\leq 1.  \label{boundpu}
\eeqa

For arbitrary time of evolution $t$, 
use of the Hubbard-Stratonovich transformation yields the following integral expression for the purity
\begin{equation}
\label{purityt-hs}
P(t)=\sqrt{\frac{1}{8\pi \gamma t}}\int_{-\infty }^{\infty }e^{-\frac{y^{2}%
}{8\gamma t}}\left\vert \frac{Z(\beta +iy)}{Z(\beta )}\right\vert ^{2}%
\mathrm{d}y,
\end{equation}%
in terms of the analytic continuation of the partition function. The term $\left\vert \frac{Z(\beta +iy)}{Z(\beta )}\right\vert ^{2}$
has been extensively studied as a characterization of the spectral
properties of quantum chaotic systems and as a proxy for information
scrambling; see \cite{Cotler17,Dyer17,delcampo17} and references therein. 

At long-times and in the absence of degeneracies, the purity  saturates at the value
\begin{equation}
P(\infty)=\frac{1}{Z(\beta )^{2}}\sum_{k}e^{-2\beta E_{k}}=\frac{Z(2\beta )%
}{Z(\beta )^{2}}, \label{satzb}
\end{equation}%
which is precisely the purity of a canonical thermal state. This long-time
asymptotic limit is shared by the unitary dynamics \cite{Dyer17,delcampo17}.

In a general setting the spectrum may exhibit degeneracies. It is then convenient to introduce the function \cite{Dyer17}
\beqa
G(\beta)=\lim_{L\to \infty}\frac{1}{L}\int^{L/2}_{-L/2}\mathrm{d}y Z(\beta-iy)Z(\beta+iy),
\eeqa
satisfying 
\beqa
Z(2\beta)\leq G(\beta)\leq Z(\beta)^2.
\eeqa
In terms of it, the long-time asymptotics of the purity reads
\begin{equation}
\label{longpure}
P(\infty)=\frac{G(\beta )%
}{Z(\beta )^{2}}.
\end{equation}%

Thus, as time goes by, the purity decays  monotonically from unit value at $t=0$ approaching the asymptotic bound (\ref{longpure}).

\section{Quantum R\'enyi Entropy}
The preceding characterization of the purity can be extended to the family of  quantum R\'enyi entropies,
which encode additional information as a function of the parameter $\alpha$.
The quantum R\'enyi entropy is defined as
\begin{equation}
S_{\alpha}(\rho)=\frac{1}{1-\alpha}\log \tr\rho^\alpha, \quad \alpha\geq 0.
\end{equation}%

We are interested in its use to characterize the evolution of the TFD under decoherence.
For an integer value $\alpha=n$, a replica calculation yields
\begin{equation}
\tr[\rho(t)^n]=\frac{1}{Z(\beta)^n}\sum_{\{k_i\}_{i=1}^n}e^{-\beta\sum_{i=1}^nE_{k_i}-\gamma t\sum_{i=1}^n(E_{k_i}-E_{k_{i+1}})^2},
\end{equation}%
which naturally reduces for $n=2$ to the purity $P(t)$, discussed in the previous section.

It is obvious that the following inequality  holds
\ba
\frac{Z(n\beta)}{Z(\beta)^n}\leq \mbox{Tr}[\rho(t)^n]\leq 1 .  \label{ineqp}
\ea

Using the Hubbard-Stratonovich transformation, one finds for arbitrary time of evolution
\begin{eqnarray}
\label{trrhon}
\tr[\rho(t)^n]&=&\frac{1}{Z(\beta)^n}\left(\frac{1}{4\pi \gamma t}\right)^\frac{n}{2}\int\prod_{i=1}^{n}\mathrm{d}y_ie^{-\sum_i\frac{y_i^{2}}{4\gamma t}}
 \sum_{\{k_i\}_{i=1}^n}e^{-\beta\sum_{i=1}^nE_{k_i}-iy_i(E_{k_i}-E_{k_{i+1}})}\\
&=&\frac{1}{Z(\beta)^n}\left(\frac{1}{4\pi \gamma t}\right)^\frac{n}{2}\int\prod_{i=1}^{n}\mathrm{d}y_ie^{-\sum_i\frac{y_i^{2}}{4\gamma t}}\prod_{i=1}^nZ(\beta+i(y_i-y_{i-1})).
\end{eqnarray}

The $n$-th quantum R\'enyi entropy reads 
\beqa
S_{n}[\rho(t)]=\frac{1}{1-n}\log \sum_{\{k_i\}_{i=1}^n}e^{-\beta\sum_{i=1}^nE_{k_i}-\gamma t\sum_{i=1}^n(E_{k_i}-E_{k_{i+1}})^2}-\frac{n}{1-n}\log Z.
\eeqa
Clearly, $\dot{S}_{n}[\rho(t)]\geq 0$, indicating that quantum Renyi entropy grows monotonically as a function of time as a result of noise-induced decoherence.

We note that at short-times
\begin{equation}
\label{tr_rhon_short}
\tr[\rho(t)^n]=1-2n\gamma t \mathrm{var}_{\rho_{\beta }}(H)+\mathcal{O}(t^2)=1-\frac{n}{2}D+\mathcal{O}(t^2),
\end{equation}%
where  we identify the decoherence rate $D$, defined in terms of the purity according to Eq. (\ref{sapurity}). 
By contrast, one finds that at long times, assuming the absence of degeneracy of energy spectrum,
\beqa
\tr[\rho(t)^n]\rightarrow\tr[\rho(\infty)^n]= \frac{Z(n\beta)}{Z(\beta)^n}.
\eeqa
This leads to the relation
\ba
\lim_{L_1,L_2,\ddd, L_n\to\infty}\prod_{i=1}^{n}\frac{1}{L_i}\int^{L_i/2}_{-L_i/2}\mathrm{d}y_i Z(\beta+i(y_i-y_{i-1}))=Z(n\beta).
\ea

In principle, using these results we can describe the behavior of the von Neumann entropy
\begin{equation}
S(\rho)=-\tr{\rho\log\rho}=-\frac{\partial}{\partial n} \tr\rho^n\big|_{n=1}=-\frac{\partial}{\partial n}\log \tr\rho^n\big|_{n=1}.
\end{equation}%
However,  the short-time expansion Eq. (\ref{tr_rhon_short}) breaks down in this limit.

As for the late-time behavior, the asymptotic density matrix is given by $\rho(\infty)$ in Eq. (\ref{rholongt}) and its von Neumann entropy equals the thermodynamic entropy
\begin{equation}
S(\rho)=\beta\langle H\rangle_{\rm \beta}+\log Z=\beta (\langle H\rangle_{\rm \beta}-F),
\end{equation}
where $F$ is the free energy.
This is most easily seen by noticing that
\beqa
S[\rho(\infty)]=-\frac{\partial}{\partial n}\log\frac{Z(n\beta)}{Z(\beta)^n}\bigg|_{n=1}=-\frac{\partial}{\partial n}\log\tr[e^{-n\beta}]\big|_{n=1}+\log Z.
\eeqa

\section{Logarithmic Negativity}
Decoherence is expected to suppress quantum correlations in the thermofield double state. To describe the evolution of entanglement, we focus on   
the logarithmic negativity, that is a convenient entanglement monotone introduced in \cite{Plenio05}. Its study in CFTs has been advanced in 
\cite{CCT12}. The explicit definition of the logarithmic negativity is
\beqa
\mathcal{E}=\log\tr|\rho_{PT}|_1=\lim_{n\rightarrow1/2}\log\tr(\rho_{PT}^{2n}),
\eeqa
where the trace norm is defined as $|A|_1=\tr\sqrt{AA^\dag}$ and $\rho_{PT}$ denotes the partial transpose of $\rho$ with respect to, e.g., the right copy.

For its computation, we note that the partial transpose of the time-evolving density matrix (\ref{eq:SMevolution}) reads 
\beqa
\rho_{PT}=\frac{1}{Z(\beta )}\sum_{k,\ell }e^{-\frac{\beta }{2}(E_{k}+E_{\ell
})}e^{-i\frac{2t}{\hbar }(E_{k}-E_{\ell })-\gamma t(E_{k}-E_{\ell
})^{2}}|k\rangle |\ell\rangle \langle \ell |\langle k |.
\eeqa
Using it, it follows that
\beqa
\log\tr(\rho_{PT}^{2n})
=\log\left[\frac{1}{Z(\beta)^{2n}}\sum_{k\ell}e^{-n\beta(E_k+E_\ell)-2n\gamma t(E_k-E_\ell)^2}\right],
\eeqa
and taking the limit $n\rightarrow 1/2$,  one finds the logarithmic negativity explicitly,
\beqa
\mathcal{E}(t)&=&\log\left(\frac{1}{Z(\beta)}\sum_{k\ell}e^{-\frac{\beta}{2}(E_k+E_\ell)-\gamma t(E_k-E_\ell)^2}\right)\\
&=&\log\left(\sqrt{\frac{1}{4\pi \gamma t}}\int_{-\infty }^{\infty }\mathrm{d}ye^{-\frac{y^{2}%
}{4\gamma t}}\frac{\left\vert Z(\beta/2 +iy) \right\vert ^{2}}{Z(\beta)}\right).
\eeqa
For the TFD (pure) state at time $t=0$,
\beqa
\mathcal{E}(0)=\log\frac{Z(\beta/2)^2}{Z(\beta)}=\beta[F(\beta)-F(\beta/2)], 
\eeqa
where $F(\beta)=-\frac{1}{\beta}\log Z(\beta)$ is the free energy.
This is equal to the $1/2$-R\'enyi entropy, as expected.

Under decoherence, 
\beqa
\dot{\mathcal{E}}(0)=-4\gamma\frac{d^2}{d\beta^2}Z(\beta/2)=-\frac{1}{4}D(\beta/2),
\eeqa
i.e., the (short-time) decay rate of the logarithmic negativity is related to (one fourth of) the decoherence time of a thermal state at temperature $\beta/2$.

At an arbitrary time of evolution $t$, the rate of change is negative  
\beqa 
\dot{\mathcal{E}}(t)
&=&-\gamma\frac{\sum_{k\ell}(E_k-E_\ell)^2e^{-\frac{\beta}{2}(E_k+E_\ell)-\gamma t(E_k-E_\ell)^2}}{\sum_{k\ell}e^{-\frac{\beta}{2}(E_k+E_\ell)-\gamma t(E_k-E_\ell)^2}}<0.
\eeqa
Thus, the logarithmic negativity  decreases monotonically to the asymptotic value
\beqa
\mathcal{E}(\infty)=\log\frac{G(\beta/2)}{Z(\beta)}\leq \mathcal{E}(0). 
\eeqa

\section{Case Studies}

\subsection{CFTs with no degenerated energy eigenvalues}

To study properties of decoherence for generic CFTs, we assume that there is no degeneracy in the 
energy spectrum.  In the absence of degeneracies in the energy spectrum,  we find in the late time limit $t\to\infty$
\ba
\mbox{Tr}[\rho(t=\infty)^n]= \frac{Z(n\beta)}{Z(\beta)^n}.
\ea
Thus, in this case  the late-time R\'enyi entropy is given by 
\ba
S^{(n)}(t=\infty)=\frac{1}{1-n}\log\frac{Z(n\beta)}{Z(\beta)^n}.
\ea
In the von-Neumann entropy limit $n\to 1$ we obtain 
\ba
S^{(1)}(t=\infty)=S_{th}(\beta),
\ea
which is the original thermal entropy, i.e., the entanglement entropy between the CFT$_1$ and CFT$_2$ 
for the TFD state  (\ref{TFD}) at $t=0$. It follows that the original entanglement entropy between CFT$_1$ and CFT$_2$
is converted into the entanglement entropy between CFT$_{1,2}$  and the environment, as expected from the 
decoherence.

For $n=2$ we find 
\beqa
G(\beta)=Z(2\beta).
\eeqa
Thus, we have 
\beqa
{\cal{ E}}(t=\infty)=\log\frac{G(\beta/2)}{Z(\beta)}=0,
\eeqa
which confirms the vanishing entanglement between the two CFTs 
in the late time limit as expected.  
According to the EPR$=$ER conjecture, this corresponds to the closing of the Einstein-Rosen bridge in the gravitational dual. 
We shall revisit this observation in Section \ref{SecTPF} where we analyze the evolution of two point functions.

\subsection{Decoherence of two-dimensional Dirac fermions}

Next, as a special class of CFTs, that are characterized by integrability, we would like to study the behavior of decoherence in a free fermion CFT, where the energy spectrum has a degeneracy.
Two dimensional CFTs are conveniently studied on a torus. The moduli of the torus is parameterized by $\tau$.
The standard definition of the partition function in statistical mechanics is then promoted to
\beqa
Z(\tau,\bar{\tau})=\tr \left(e^{-{\rm Im}\tau H+i{\rm Re}\tau P}\right),
\eeqa
where the Hamiltonian and the momentum operator in a cylinder of width $L$ are given by
\beqa
H=\frac{2\pi}{L}\left(L_0+\bar{L}_0-\frac{c}{12}\right), \qquad P=\frac{2\pi}{L}\left(L_0-\bar{L}_0\right),
\eeqa
in terms of the Virasoro generators $L_0$ and $\bar{L}_0$. 
The decoherence mechanism leading to the master equation (\ref{ME}) and the evolution of the density matrix (\ref{eq:SMevolution}) stems from fluctuations in the Hamiltonian only.
For consistency, we shall focus on ${\rm Re}\tau=0$.

The partition function of two-dimensional Dirac fermions can be written  as \cite{Azeyanagi08}
\begin{equation}
Z(\beta)=\frac{|\theta_3(0|\tau)|^2}{|\eta(\tau)|^2}, 
\end{equation}%
where  the  Jacobi theta function reads $\theta_3(z,\tau)=\sum_{n=-\infty}^\infty\exp(i\pi n^2\tau+i2\pi n z)$ and the Dedenkin function 
is given by $\eta(\tau)=e^{i\pi\tau/12}\prod_{m=1}^\infty(1-e^{i2\pi  m\tau})$. The space direction is compactified as $x\sim x+L$  and $\tau=i\beta/L$.
The explicit form  of the partition function can be written as an infinite product  
\begin{eqnarray}
Z(\beta)&=&e^{\frac{\pi}{6}\beta L}\prod_{m=1}^\infty\left(1+e^{-2\pi\frac{\beta}{L}(m-1/2)}\right)^4\label{ZbCFT}\\
&=&e^{\frac{\pi L}{6\beta}}\prod_{m=1}^\infty\left(1+e^{-\frac{2\pi L}{\beta}(m-1/2)}\right)^4=Z(L^2/\beta),
\end{eqnarray}
where the last line is a consequence of the modular invariance $Z(\beta)=Z(L^2/\beta)$.
We find the following behavior in the high temperature limit $\beta\rightarrow 0$, 
\beqa
Z(\beta)\sim e^{\frac{\pi}{6}\beta L}.
\eeqa

Expression (\ref{ZbCFT}) for the partition function is particularly convenient to determine the decoherence rate characterizing the short-time asymptotics of the purity. One obtains
\beqa
\label{DCFTexact}
D=4\gamma \frac{\mathrm{d}^{2}}{\mathrm{d}\beta ^{2}}\ln \left[ Z(\beta )\right] 
=\frac{4\gamma \pi^2}{L^2}
\sum_{m=1}^\infty\frac{(1 - 2 m)^2}{{\rm cosh}^2[\beta \pi(m - 1/2)]}.
\eeqa
The rate is thus divergent in the high-temperature limit,
\beqa
\label{DCFThigh}
D\simeq \frac{4\pi}{3}\gamma\frac{L}{\beta^3}.
\eeqa
By contrast, in the low temperature case 
$\beta\gg L$, it is well described by 
\beqa
\label{DCFTlow}
D=16 \pi^2 \frac{\gamma}{L^2}e^{-\pi\frac{\beta}{L}}.
\eeqa

\begin{figure}[tbp]
\centering{}\includegraphics[width=3.15in]{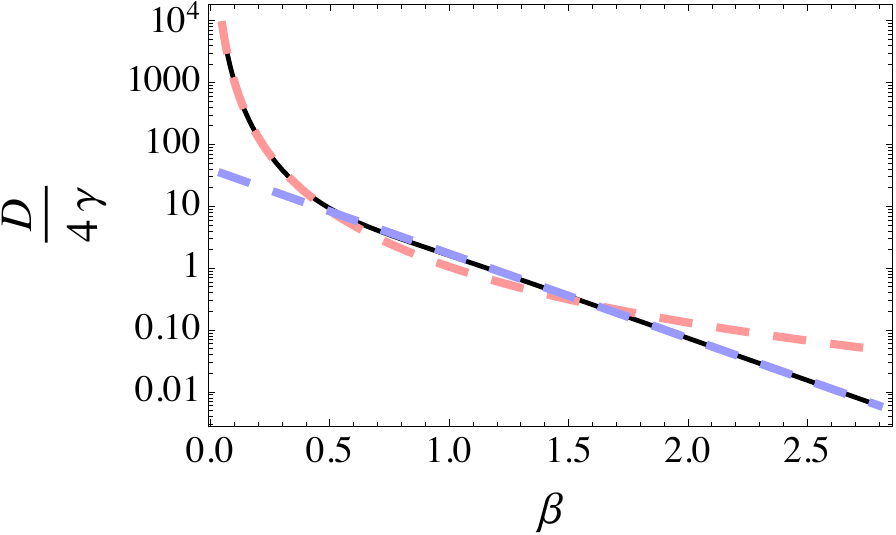}
\caption{Decoherence rate of two-dimensional Dirac fermions. 
The numerically-exact decoherence rate  evaluated by truncation of Eq. (\ref{DCFTexact}), shown as a solid black line,  is compared to the high and low temperature asymptotes in Eqs. (\ref{DCFThigh}) and (\ref{DCFTlow}) that are shown in light red and blue color, respectively.
}
\label{fig1}
\end{figure}
For arbitrary times of evolution, the purity can be computed using its integral representation (\ref{purityt-hs}),
where the spectral form factor simplifies to
\begin{equation}
f(\beta,y)=\left\vert \frac{Z(\beta +iy)}{Z(\beta )}\right\vert^{2}=
\prod_{m=1}^\infty\left(\frac{\cos[2\pi y(m-1/2)/L]+\cosh[2\pi y(m-1/2)/L]}{1+\cosh[2\pi \beta(m-1/2)/L]}\right)^4
\leq 1.
\end{equation}

\begin{figure}[tbp]
\centering{}\includegraphics[width=3.15in]{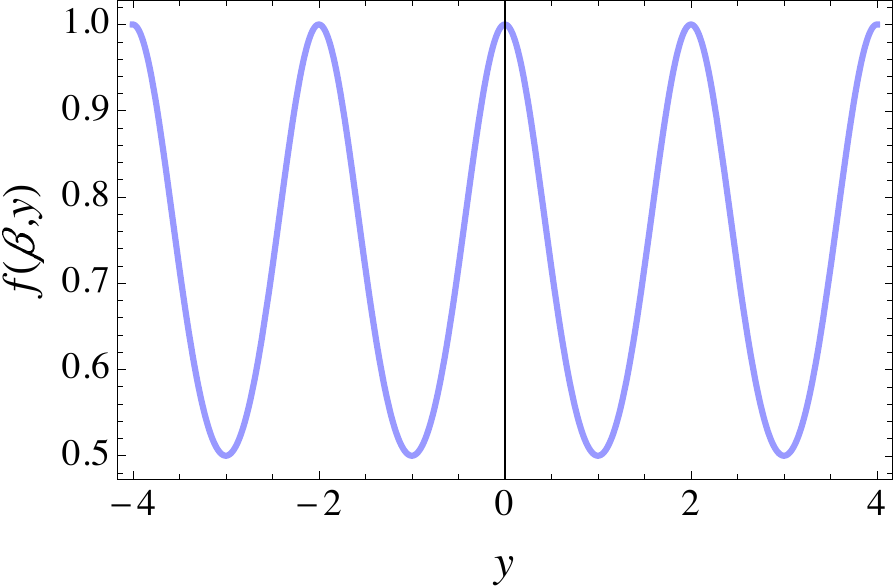}
\centering{}\includegraphics[width=3.15in]{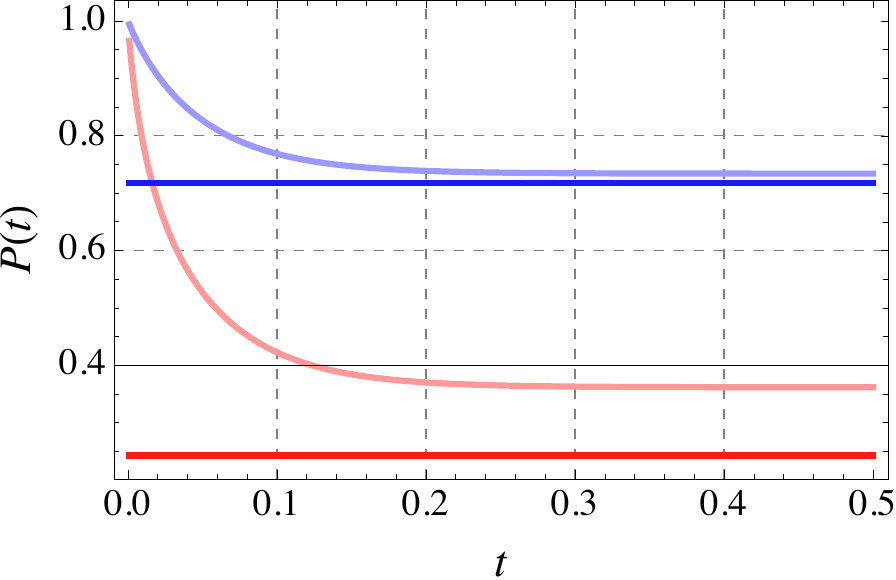}
\caption{Decoherence of two-dimensional Dirac fermions. 
(a) Periodicity of the function $f(\beta,y)$ as a function of the time of evolution $t$. 
(b) Time evolution of the purity for a thermofield double state prepared at $t=0$ under decoherence, when the evolution is governed by  the master equation (\ref{ME}). 
The asymptotic value of the purity exceeds the value $Z(2\beta)/Z(\beta)^2$, indicating the presence of degeneracies in the energy spectrum. Cases with $\beta=1/2,1$ with corresponding values $Z(2\beta)/Z(\beta)^2=0.243, 0.718$ are shown in red and blue color, respectively.}
\label{fig2}
\end{figure}

In the time domain, the function $f(\beta,y)$ is even $f(\beta,y)=f(\beta,-y)$ and periodic, satisfying $f(\beta,y)=f(\beta,y+2n)$ for all integer $n\in\mathbb{Z}$, as shown in Fig. \ref{fig2}(a). 
In this case we can rewrite the purity $P(t)=\mbox{Tr}[\rho(t)^2]$ as follows
\ba
P(t)&=&\frac{1}{\s{8\pi\gamma t}}\int^\infty_{-\infty}\mathrm{d}y e^{-\frac{y^2}{8\pi\gamma t}}f(\beta,y) \no
&=&\frac{1}{\s{8\pi\gamma t}}\sum_{n=-\infty}^{\infty}\int^1_{-1}\mathrm{d}y e^{-\frac{(y+2n)^2}{8\pi\gamma t}}f(\beta,y) \no
&=&\frac{1}{2}\sum_{m=-\infty}^\infty \int^1_{-1}\mathrm{d}y e^{-2 \pi^2\gamma t m^2+\pi i y m}f(\beta,y),  \label{www}
\ea
where in the final line we employed the Poisson resummation formula
\ba
\sum_{n=-\infty}^\infty e^{-\frac{\pi(n-b)^2}{a}}=\s{a}\sum_{m=-\infty}^\infty e^{-\pi a m^2+2\pi i bm}.
\ea

At the late time limit $t\to\infty$ we simply obtain
\ba
\lim_{t\to\infty}\mbox{Tr}[\rho(t)^2]= \int^1_0 \mathrm{d}y f(\beta,y)\leq 1,  \label{final}
\ea
and numerically we can confirm  that
\ba
\int^1_0 \mathrm{d}y Z(\beta-iy)Z(\beta+iy)> Z(2\beta).  \label{ggg}
\ea
This leads to 
\ba
\lim_{t\to\infty}\mbox{Tr}[\rho(t)^2]> \frac{Z(2\beta)}{Z(\beta)^2}.
\ea
For example, when $\beta\gg1 $ we find 
\ba
\frac{Z(\beta-iy)Z(\beta+iy)}{Z(2\beta)}\simeq 1+8\cos(\pi y/L)e^{-\pi\frac{\beta}{L}}+12\cos^2(\pi y/L)e^{-2\pi\frac{\beta}{L}},
\ea
which obviously leads to (\ref{ggg}).  This means that the lowest bound of the purity (\ref{boundpu}) is not saturated and this is due to the degeneracy of energy eigenvalues as we will see below.

The full  decay dynamics of the purity   as a function of time is shown in Fig. \ref{fig2}(b).  As already mentioned, the monotonic decay is a feature that holds for any state  evolving under the master equation (\ref{ME}), 
which can be associated with a unital map, and thus satisfies the conditions for strictly purity-decreasing quantum Markovian dynamics \cite{Lidar06}.
Regarding its late-time behavior, we note that 
as energy levels $E_m$ may have a degeneracy $h_m$, the purity approaches asymptotically the value
\begin{equation}
P(\infty)=\frac{\sum_mh_m^2e^{-2\beta E_m}}{Z(\beta)^2}>\frac{Z(2\beta)}{Z(\beta)^2},
\end{equation}
where $Z(\beta)=\sum_mh_me^{-\beta E_m}$, in terms of the energy eigenstates.
Similarly, using the replica trick we find
\begin{equation}
\tr[\rho(\infty)^n]=\sum_m\left(\frac{h_me^{-\beta E_m}}{Z(\beta)}\right)^n=\sum_m\lambda_m^n,
\end{equation}
where $\lambda_m$ are the eigenvalues of $\rho(\infty)$.
The corresponding von Neumann entropy
\begin{eqnarray}
S[\rho(\infty)]&=&-\frac{\partial}{\partial n}\tr[\rho(\infty)^n]=-\frac{\partial}{\partial n}\sum_m\left(\frac{h_me^{-\beta E_m}}{Z(\beta)}\right)^n\bigg|_{n=1}\nonumber\\
&=&-\sum_m\left(\frac{h_me^{-\beta E_m}}{Z(\beta)}\right)\log\left(\frac{h_me^{-\beta E_m}}{Z(\beta)}\right),
\end{eqnarray}
which matches the thermal value  associated with  a canonical Gibbs  state with degeneracies in the energy spectrum.

Finally, the logarithmic negativity in the late time limit is found to be non-vanishing 
\ba
{\cal{ E}}(t=\infty)=\log \frac{G(\beta/2)}{Z(\beta)}=\log \frac{Z(\beta/2)^2\int^1_{0}\mathrm{d}y f(\beta/2,y)}{Z(\beta)}>0.
\ea
This manifestly shows that a non-zero quantum entanglement between the two CFTs  remains in the late time limit.

\subsection{Decoherence in holographic CFTs}

To study the decoherence in the maximally chaotic CFTs, we would like to study the holographic CFTs in two dimensions.  Consider the case in which the Hamiltonian $H$ is associated with a 2d holographic CFT on a circle (with periodicity $2\pi$ such that $x\sim x+2\pi$).
In the 2d CFTs dual to classical gravity, the partition function in the high temperature phase  (BTZ black hole) reads 
\beqa
Z(\beta)\sim e^{\frac{\pi^2 c}{3\beta}}.  \label{btz}
\eeqa 

From the equilibrium value of the thermal energy variance \begin{equation}
\mathrm{var}_{\rho _{\beta }}(H)=\frac{\mathrm{d}^2}{\mathrm{d}\beta^2}\log Z(\beta)=\frac{2\pi^2 c}{3\beta ^3},
\end{equation}
we can estimate the  decoherence time  to be
\beqa
\label{DHolo}
D=\frac{8\pi^2}{3}\frac{\gamma c}{\beta^3}.
\eeqa
This decoherence rate governs the short-time asymptotics of the purity, i.e., according to the expansion in Eq. (\ref{sapurity}).

In the long-time limit, since we do not expect degeneracy of energy levels in chaotic CFTs\footnote{
Note that the non-renormalization property due to supersymmetries can give a degeneracy of energy levels.
However, the contributions of supersymmetric states (BPS states) to the partition function is much smaller compared with the non-supersymmetric states. Thus we ignore such contributions.}
, we expect 
\beqa
\label{PHoloInf}
P(\infty)=\lim_{t\to\infty}\mbox{Tr}[\rho(t)^2]=\frac{Z(2\beta)}{Z(\beta)^2}\sim \exp\left[-\frac{\pi^2c}{2\beta}\right].
\eeqa
Note that this asymptotic value of the purity decays from unit value exponentially as a function of $c/\beta$.
This predicts 
\beqa
G(\beta)=\int^{\infty}_{-\infty}\mathrm{d}y Z(\beta-iy)Z(\beta+iy)\sim e^{\frac{\pi^2 c}{6\beta}}.
\eeqa

To describe the decohering dynamics at intermediate times, we estimate the purity via the integral representation in terms of the analytically continued partition function
\beqa
P(t)&=&\sqrt{\frac{1}{8\pi \gamma t}}\int_{-\infty }^{\infty }\mathrm{d}ye^{-\frac{y^{2}%
}{8\gamma t}}\left\vert \frac{Z(\beta +iy)}{Z(\beta )}\right\vert ^{2}%
\\
&=&\sqrt{\frac{1}{8\pi \gamma t}}e^{-\frac{2\pi^2 c}{3\beta}}\int_{-\infty }^{\infty }\mathrm{d}ye^{-\frac{y^{2}%
}{8\gamma t}}e^{\frac{2\pi^2c}{3}\frac{\beta}{y^2+\beta^2}}.%
\eeqa

\begin{figure}[tbp]
\centering{}\includegraphics[width=3.15in]{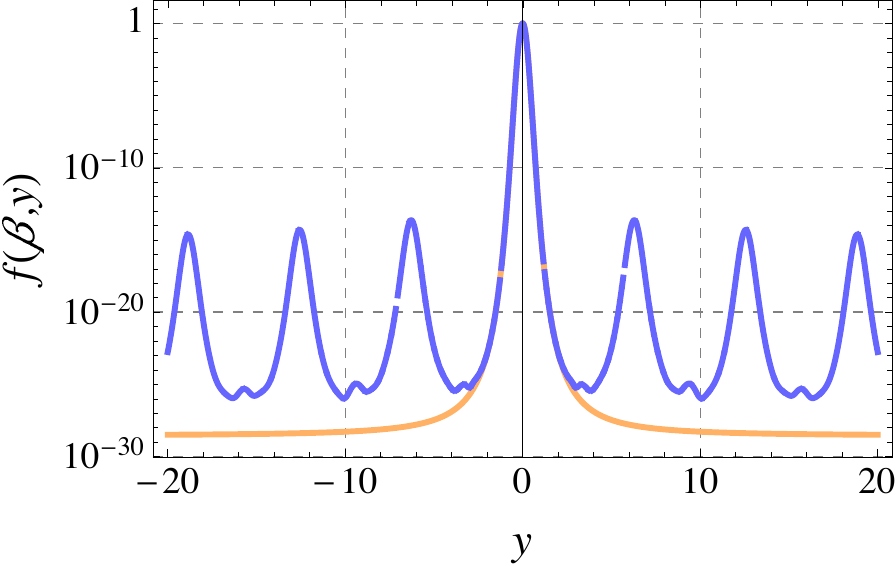}
\caption{
Spectral form factor $f(\beta,y)$ for the holographic CFT. In the numerically exact result (in blue) using (\ref{Zgrav}), the contribution of the sum over saddles is subdominant and orders of magnitude smaller than that of the leading one (orange).
}
\label{fig3}
\end{figure}

The preceding analysis relied on the use of the classical gravity result (i.e. BTZ black hole) for the partition function in terms of the leading saddle.
Consider now going beyond this approximation. 
The correct partition function is given by the summation over all saddles (i.e. classical solutions to Einstein equations) \cite{Strominger98,Dyer17}:
\beqa
\label{Zgrav}
Z(\beta-iy)=Z_{gravity}(\tau,\bar{\tau})=\sum_{(p,q,r,s)\in\mathbb{Z},ps-qr=1}\exp\left[-i\frac{\pi c}{12}\left(\frac{p\tau+q}{r\tau+s}-\frac{p\bar{\tau}+q}{r\bar{\tau}+s}\right)\right], 
\eeqa
where the moduli parameter read
\beqa
\tau=i\left(\frac{\beta-iy}{2\pi}\right),\quad \bar{\tau}=-i\left(\frac{\beta-iy}{2\pi}\right).
\eeqa
The expression for $Z(\beta+iy)$ follows by replacing $y\rightarrow-y$. 

The expression (\ref{Zgrav}) 
takes into account only the leading classical gravity contributions from all non-perturbative saddles (or equally $SL(2,\mathbb{Z})$ summations).
Therefore we ignore loop corrections in the gravity. However, we know that at each saddle, 
 the classical gravity contribution dominates over its loop correction and thus the summations of all classical 
saddles are crucial. In this sense, we expect that the essential feature of the full gravity partition function for our purpose can be captured by the above  (\ref{Zgrav}), neglecting loop corrections. 

The integral representation of the purity involves the integrand $f(\beta,y)=\left\vert \frac{Z(\beta +iy)}{Z(\beta )}\right\vert ^{2}$ that is depicted in Figure \ref{fig3}, by truncating the sum over $SL(2,\mathbb{Z})$  and checking for convergence. Specifically, by restricting $p,q,r,s\leq  n_{\rm max}$, taking $n_{\rm max}=10$ we find 1012 tuples $(p,q,r,s)$ satisfying $ps-qr=1$. 
The contribution of higher-order saddles is clearly subdominant, being many orders of magnitude below the leading saddle, even in the absence of decoherence.

The emergence of classical gravity result associated with the leading saddle is further guaranteed by decoherence as the peaks away of the origin are suppressed by the Gaussian factor. Yet, as the time of evolution goes by, this factor becomes broader and broader.

We are interested in the function
\beqa
G(\beta)&=&\int_{-\infty }^{\infty }\mathrm{d}yZ(\beta-iy)Z(\beta+iy)\\
&=&\int_{-\infty }^{\infty }\mathrm{d}y\left(\sum_{(p,q,r,s)\in\mathbb{Z},ps-qr=1}e^{\frac{\pi^2 c}{3}\frac{\beta-iy}{4\pi^2s^2+(\beta-iy)r^2}}\right)\left(\sum_{(p,q,r,s)\in\mathbb{Z},ps-qr=1}e^{\frac{\pi^2 c}{3}\frac{\beta+iy}{4\pi^2s^2+(\beta+iy)r^2}}\right).\nonumber
\eeqa
To leading order, the absence of degeneracy suggests
\beqa
G(\beta)=\int_{-\infty }^{\infty }\mathrm{d}yZ(\beta-iy)Z(\beta+iy)\sim e^{\frac{\pi^2c}{6\beta}}.
\eeqa

\begin{figure}
  \centering
  \includegraphics[width=6cm]{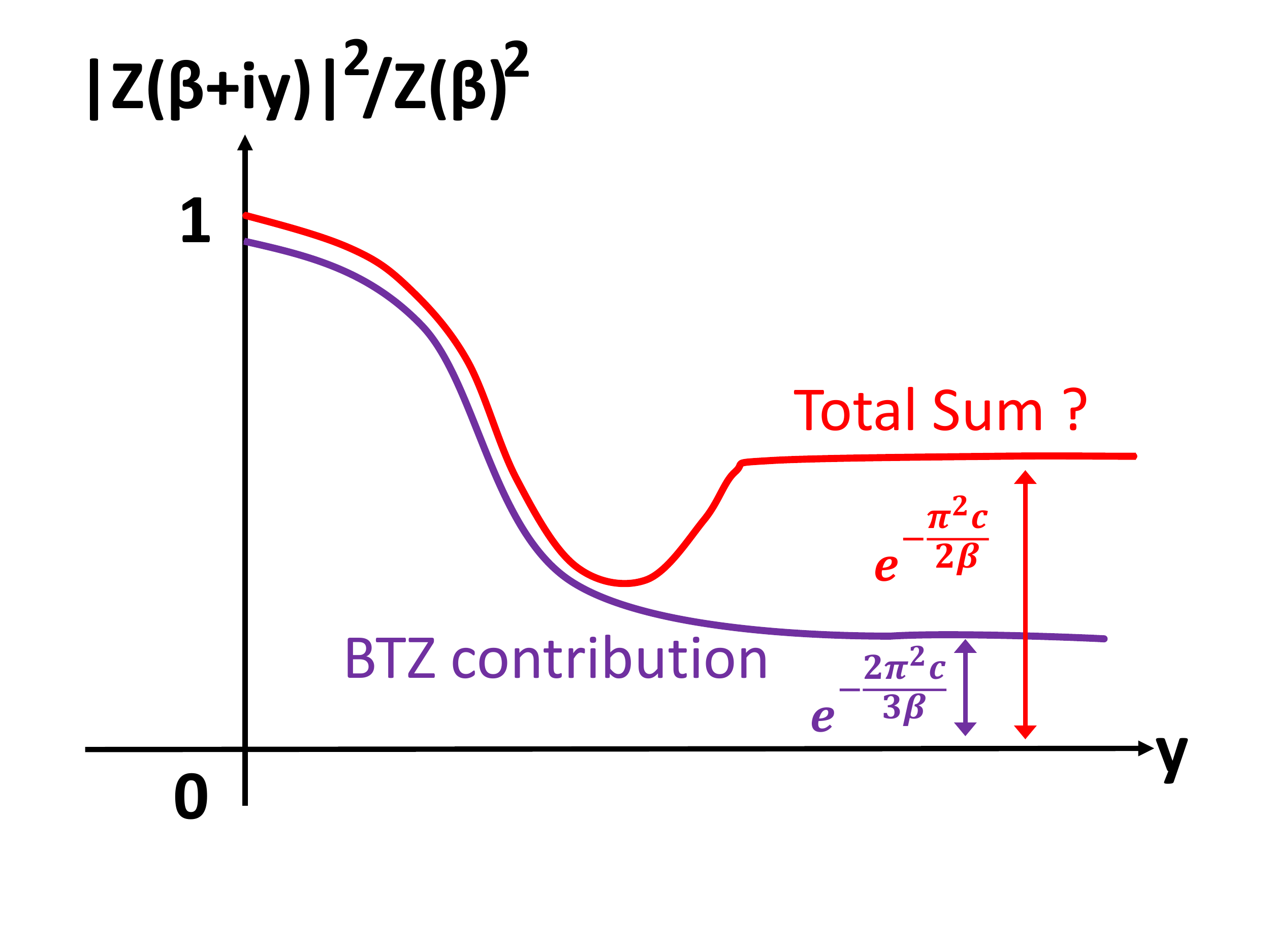}
  \caption{Behaviors of Spectrum Function $f(\beta,y)=\frac{|Z(\beta+iy)|^2}{Z(\beta)^2}$ in the high temperature phase. The purple curve describes the contribution from the BTZ black hole saddle point. The red curve is based on the speculation from the relation $P(t=\infty)= \frac{Z(2\beta)}{Z(\beta)}\sim e^{-\frac{\pi^2 c}{2\beta}}$.}
\label{spec}
\end{figure}

Let us try to speculate the behavior of the spectrum function  $f(\beta,y)=\frac{|Z(\beta+iy)|^2}{Z(\beta)^2}$ 
for holographic CFTs from what we know.  If we assume the high temperature phase $\beta<2\pi$, then the classical 
gravity dual is given by BTZ black hole before we add the decoherence. The BTZ solution leads to the spectrum function 
given from (\ref{btz}) by  
\beqa
f(\beta,y)_{BTZ}\sim e^{-\frac{2\pi^2c}{3\beta}}\cdot e^{\frac{2\beta}{\beta^2+y^2}},
\eeqa
which is depicted as the purple curve in Fig. \ref{spec}. This leads to the late time behavior of purity
\beqa
P(t=\infty)\simeq \lim_{L\to \infty} \frac{1}{L}\int^{L/2}_{-L/2}\mathrm{d}y
\frac{Z(\beta-iy)Z(\beta+iy)}{Z(\beta)^2}\sim e^{-\frac{2\pi^2c}{3\beta}},  \label{BTZS}
\eeqa

On the other hand, in the absence of degeneracy of energy levels, we expect the purity at late time $P(\infty)$, which should be obtained after the full summation over all saddle points, to look like 
\beqa
P(t=\infty)=\frac{Z(2\beta)}{Z(\beta)}\sim e^{-\frac{\pi^2 c}{2\beta}}(\gg e^{-\frac{2\pi^2c}{3\beta}}).
\label{pthol}
\eeqa
This is much larger than the BTZ saddle point result (\ref{BTZS}). To reproduce the correct result 
(\ref{pthol}), the spectrum function should behave like 
\beqa
f(\beta,y)\sim e^{-\frac{\pi^2 c}{2\beta}},
\eeqa
in the limit $y\to\infty$. It is natural to believe that this enhancement is due to the contributions from other saddle points other than the BTZ. Indeed, the late time behavior largely depends on fine-grained energy level structures which are difficult to see from the semi-classical gravity approximation.

It is also useful to study the low temperature phase $\beta>2\pi$. In this case, the dominant saddle is the thermal AdS solution if we ignore the decoherence. This simply leads to  the spectrum function 
\beqa
f(\beta,y)_{TAdS}=1.
\eeqa
Thus, the purity $P(t)$ remains unity at any time if we take into account this thermal AdS saddle:
\beqa
P(t)=1.
\eeqa
In other words, the state is not affected by  decoherence. This is consistent with (\ref{satzb}) because in the low temperature phase we have $Z(\beta)\simeq e^{\frac{c\beta}{12}}$ and therefore 
$Z(2\beta)/Z(\beta)^2\simeq 1$. This also looks plausible because the original TFD state 
at $t=0$ in the low temperature phase is dual to two copies of thermal AdS solutions which are entangled with each other only via $O(1)$ entropy. Thus the state is well approximated by the vacuum and thus the state is not affected by the 
decoherence, which gives a random factor in front of the total Hamiltonian.

\section{Two-Point functions and Gravity Dual Interpretation}\label{SecTPF}

We would like to calculate the two point functions $\la O_LO_R\lb$ of two identical primary operators $O_L$ and $O_R$, located in CFT$_L$ (the first CFT) and CFT$_R$ (the second CFT) of the thermofield double, respectively. We write their conformal dimension as $\Delta_O=h_O+\bar{h}_O$. The behavior of two point functions will reveal the geometry of gravity duals of the decoherence process in holographic CFTs.

We know that the two-point function in the TFD state  is equal to the two point function on a cylinder with  periodicity $\beta$, 
which is given by
\ba
\la {\rm TFD}|e^{it_L H_L-it_R H_R}O_L O_R e^{-it_L H_L+it_R H_R}|{\rm TFD}\lb=\left(\frac{1}{\frac{\beta}{\pi}\cosh\left[\frac{\pi(t_L-t_R)}{\beta}\right]}\right)^{2\Delta_O},  \label{twopta}
\ea
where we assumed that $O_L$ and $O_R$ are inserted in the same location with respect to the spatial coordinate $x$. Notice that the time $t_L$ and $t_R$ flow oppositely such that the time evolution of the thermofield double  state is given by $e^{-it_L H_L+it_R H_R}|{\rm TFD}\lb$. For example, the time evolution $t_L=t_R$ is trivial and does not change the thermofield double state.

Thus we find the two point function for our state $\rho(t)$:
\ba
\la O_LO_R\lb_{\rm DEC}&=&\mbox{Tr}[\rho(t)O_LO_R]  \no
&=&\frac{1}{\s{4\pi\gamma t}}\int^\infty_{-\infty}\mathrm{d}y e^{-\frac{y^2}{4\gamma t}}\left(\frac{1}{\frac{\beta}{\pi}\cosh\left[\frac{2\pi(t-y/2)}{\beta}\right]}\right)^{2\Delta_O}.
\ea

At $t=0$, this reproduces the known result for the TFD state $\la O_LO_R\lb=\left(\frac{\pi}{\beta}\right)^{2\Delta_O}$ at $t=0$.  We are interested in the behavior in the limit $t\to\infty$. It is clear that for the TFD state we have
\ba
\la O_LO_R\lb_{\rm TFD} \sim e^{-4\pi\Delta_O t/\beta},  \label{TFDtwo}
\ea
in the late time limit as we obtain from (\ref{twopta}) by setting $t_L=-t_R=t$.

Let us assume $\Delta_O$ is very large so that we can apply the saddle point approximation $y=y_*$ of the integral 
$\int dy$. We assume $t-y_*/2>0$ at the saddle point, as it is easy to see that $t -y_*/2<0$ is incompatible
with the saddle point equation.  By approximating 
$\cosh\left[\frac{2\pi(t-y/2)}{\beta}\right]\sim e^{\frac{2\pi}{\beta}(t-y/2)}$, the saddle point equation reads
\ba
\frac{\mathrm{d}}{\mathrm{d}y}\left[-\frac{y^2}{4\gamma t}-\frac{4\pi \Delta_O}{\beta}(t-y/2) \right]=0.
\ea
The saddle point is solved as 
\ba
y_*=\frac{4\pi\gamma\Delta_O}{\beta}t,
\ea
and the integral is approximated as 
\ba
\la O_LO_R\lb_{\rm DEC}\sim e^{-4\pi\Delta_O t/\beta} e^{\frac{4\pi^2\Delta_O^2\gamma}{\beta^2}t},
\ea

However, this is consistent with the assumption $t-y_*/2>0$ only if $\frac{2\pi\gamma\Delta_O}{\beta}<1$.
When $\frac{2\pi\gamma\Delta_O}{\beta}>1$, the dominant contribution to the integral will come from the point 
$t-y_*/2=a_*$, where $a_*$ is a finite positive constant.  This leads to the estimation $ \la O_LO_R\lb_{\rm DEC}\sim e^{-t/\gamma}$.

\begin{figure}[t]
  \centering
  \includegraphics[width=3.15in]{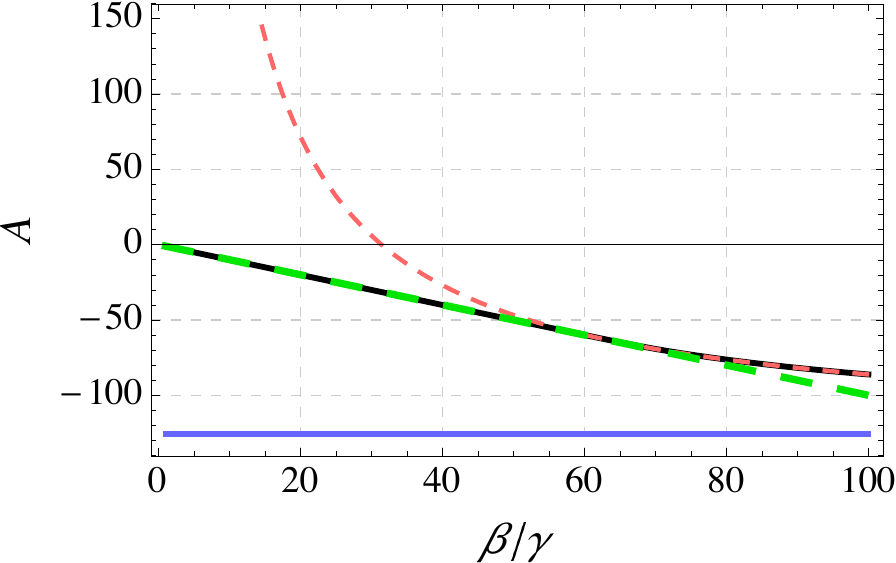}
  \caption{Plots of the coefficient $A$ in the late time behavior of the two point function $\la O_LO_R\lb \sim e^{At/\beta}$ as a function 
of $\beta/\gamma$. The black curve describes $A$ in our decohered state. The red dashed curve corresponds to the approximation given by (\ref{twoone}): 
$A=-4\pi\Delta_O+4\pi^2\Delta_O^2\gamma/\beta$.
The green dashed curve corresponds to the approximation given by (\ref{twotwo}):  $A=-\beta/\gamma$. In these plots, we set $\Delta_O=10$.
The blue line shows the value of $A$ for the TFD state (\ref{TFDtwo}).}
\label{comp}
\end{figure}

In summary, we find for large $\Delta_O$ that
\ba
&& \mbox{When}\ \frac{2\pi\gamma\Delta_O}{\beta}<1:\ \  \la O_LO_R\lb_{\rm DEC}\sim e^{-4\pi\Delta_O t/\beta} e^{\frac{4\pi^2\Delta_O^2\gamma}{\beta^2}t}, \label{twoone}
\\
&&  \mbox{When}\ \frac{2\pi\gamma\Delta_O}{\beta}>1:\ \  \la O_LO_R\lb_{\rm DEC}\sim  e^{-t/\gamma}.  \label{twotwo}
\ea
We can confirm, this behavior in the plots of Fig. \ref{comp}.

By comparing these equations with the TFD result (\ref{TFDtwo}), we find that the two point function in the decohered state is much larger than that in the TFD state. The geodesic length $\Gamma$ in the gravity dual is related to the two point function via the standard rule $\la O_LO_R\lb\sim e^{-\Delta_O \Gamma}$. Thus, we expect the geodesic length from one boundary to the other boundary in the gravity dual of the decohered state to be shorter than that in the gravity dual of TFD state (i.e., the eternal BTZ) by the amount of the two point function, which is substantially large. It seems the that the inner horizon regions in the eternal BTZ are quite squeezed in the decohered geometry. In principle, we can realize such a geometry by multiplying an overall conformal factor which is reduced only on the inner horizon regions and which takes the unit value on the outer horizon regions. It would be an intriguing future problem to work this out in detail.

However, we have to be careful given that the geodesic length $\Gamma$ is not universally defined as the two point function can have two different forms (\ref{twoone}) and (\ref{twotwo}) and is not universal. On the other hand, since the density matrix for CFT$_L$, $\rho_L=$Tr$_R[\rho(t)]$, remains the same as the original canonical distribution, the outer horizon region 
in the gravity dual is not affected by  decoherence.

 \section{Information Loss and Quantum Channel Dilations}
 
The open noisy evolution we have assumed throughout  can be thought of resulting from the interaction of the CFTs with an external environment. This interaction leads to the buildup of correlations  between the CFTs and the environment. As the degrees of freedom of the latter are unknown or inaccessible,  the entropy of the composite system of the two CFTs increases, as we have shown. This is a manifestation of  information loss in the original CFTs through its leakage to the environment. We elucidate this aspect in this section.

We first note that the noise-induced energy dephasing we have considered to introduce decoherence is just an instance of an open quantum dynamics.
A general open quantum process can be described by a quantum channel $\Lambda[\cdot]$, that is, a completely positive and trace preserving linear map  from density operators to density operators \cite{Hayashi10},
\beqa
\rho(t)=\Lambda[\rho(0)].
\eeqa
 Any quantum channel admits a Kraus decomposition 
 \beqa
 \rho(t)=\sum_{j=1}^{d_K} K_{j}\rho(0) K_{j}^\dag,
 \eeqa
in terms of a set of Kraus operators $K_{j}$ satisfying $\sum_{j} K_{j}^\dag K_{j}=\mathbf{1}$. Though different Kraus decompositions are possible, the value of $d_K$ need not be larger  than $d^2$, where $d$ is  the Hilbert space dimension of the CFT.
By Stinespring's dilation theorem, it is possible to represent a quantum channel as an isometry in an enlarged Hilbert space,
\beqa
U_\Lambda=\sum_{j=1}^{d_K}  K_{j}\otimes |j\ra_E.
\eeqa
The isometry satisfies $U_\Lambda^\dag U_\Lambda=\mathbf{1}$, with  $U_\Lambda U_\Lambda ^\dag $ being a projector onto the tensor product of the joint system.
The evolution of the latter is thus described by
\beqa
U_\Lambda \rho(0)U_\Lambda^\dag=\sum_{j\ell} K_{j}\rho(0) K_{\ell}^\dag\otimes |j\ra_E  \la \ell|_E,
\eeqa
making the interaction with the environment explicit. Tracing over the degrees of freedom of the environment recovers 
\beqa
\rho(t)=\tr_E[U_\Lambda \rho(0)U_\Lambda^\dag]=\Lambda[\rho(0)].
\eeqa

In our setting, making use of the time-dependent density matrix (\ref{rhotHS}) we note that
\ba
\rho(t)=\int^\infty_{-\infty}\mathrm{d}y K(y)|{\rm TFD}\lb\la {\rm TFD}|K(y)^\dag,  \label{decKra}
\ea
where
\ba
K(y)=\left(\frac{1}{4\pi\gamma t}\right)^{\frac{1}{4}}e^{-\frac{y^2}{8\gamma t}}e^{-i(t-y/2)(H_1+H_2)}.
\ea
This is analogous to describe the dynamics in terms of a quantum channel with aKraus decomposition in terms of Kraus operators labelled by a continuous index $y$, satisfying $\int \mathrm{d}y K(y)^\dag K(y)=1$. This does not rule out an alternative Kraus decomposition with discrete index, which indeed must exist \cite{Kraus71,Lindblad76}.
At any rate, the quantum channel can be represented by the isometry
\ba
U_\Lambda=\int^\infty_{-\infty}\mathrm{d}y K(y)\otimes |y\ra_E,
\ea
which suggests that the role of the environment is played by an infinite-dimensional system in this representation. 
The evolution in the joint system takes the form
\beqa
U_\Lambda \rho(0)U_\Lambda^\dag=\frac{1}{Z(\beta )}\left(\frac{1}{4\pi\gamma t}\right)^{\frac{1}{2}}\sum_{k,\ell }\iint \mathrm{d}y\mathrm{d}xe^{-\frac{y^2+x^2}{8\gamma t}}
e^{-\left(\frac{\beta }{2}+i\frac{2t}{\hbar }+iy\right)E_{k}}e^{-\left(\frac{\beta }{2}-i\frac{2t}{\hbar }-ix\right)E_{\ell}}|k\rangle |k\rangle \langle \ell |\langle \ell |\otimes |y\ra \la x|_E.
\eeqa
Tracing over the degrees of freedom of the environment and using the orthonormality relation for continuous variables $ \la x |y\ra=\delta(x-y)$, one recovers $\rho(t)$.

\section{Purification into GHZ-like States and Possible Gravity Duals}

As we have seen, in the absence of degeneracies in the energy spectrum expected in a holographic theory,  the long-time dynamics under decoherence gives rise to the mixed state\footnote{A holographic interpretation of the same state was discussed in \cite{Verlinde19} independently, from different view points.
Refer also to \cite{Almheiri19} for a process similar to ours in the context of information paradox.}
\begin{equation}
\rho(\infty)=\lim_{t\rightarrow\infty}\rho(t)=\frac{1}{Z(\beta )}\sum_{k}e^{-\beta E_{k}}|k\rangle
|k\rangle \langle k|\langle k|.
\label{rho2inf}
\end{equation}%
We note that the purification of this asymptotic state gives rise a generalization of the thermofield double state that involves three copies of the system
\begin{equation}
\label{tf3}
|{\rm TF}_3(\beta)\ra=\frac{1}{[Z(\beta )]^{\frac{1}{2}}}\sum_{k}e^{-\frac{\beta}{2} E_{k}} |k\rangle|k\rangle|k\rangle.
\end{equation}%
Note that this definition differs from the state introduced in \cite{Okuyama19}.  
Equation (\ref{tf3})  is indeed a purification satisfying $|{\rm TF}_3(\beta)\ra=\sqrt{\rho(\infty)}\otimes\mathbf{1}_3|\Phi\ra$ where $\sqrt{\rho(\infty)}$ is the positive semidefinite square root of the asymptotic density matrix (\ref{rho2inf}) of the two initial copies and $|\Phi\ra=\sum_k
 |k\rangle|k\rangle|k\rangle$ is the unnormalized (3-copies) maximally entangled state. Clearly, given the purified state $|{\rm TF}_3(\beta)\ra \la {\rm TF}_3(\beta)|$ and tracing over any of the three copies of the CFT leads to the recovery of the decohered state (\ref{rho2inf}). 

Importantly, only one extra copy of the CFT is needed to achieve the purification of (\ref{rho2inf}), by contrast to the standard doubling of the Hilbert space, that would require two extra copies instead. This is a consequence of the fact that even under decoherence the density matrix is at all times of the form $\rho(t)=\sum_{k\ell}\rho_{kk,\ell\ell}(t)|k\ra |k\ra \la \ell|\la\ell|$. As a result, it can be brought to a diagonal form $\rho(t)=\sum_{m}\sigma_m|m\ra |m\ra \la m|\la m|$ in a new basis of entangled states over the two CFTs initially under consideration, $|m\ra |m\ra=\sum_{k}u_{k,m}|k\ra |k\ra$ as it follows from the decomposition $\rho_{kk,\ell\ell}=\sum_mu_{k,m}\sigma_mu_{m,\ell}^*$ in terms of the unitary matrix $u$ that is $d\times d$, where $d$ is the dimension of the Hilbert space of a single copy $d={\rm dim}(\mathcal{H})$. In the final decohered state $\rho(\infty)$ it is clear that the rank (number of nonzero eigenvalues) is at most $d$, as opposed to $d^2$, expected for a generic density matrix of a bipartite system.

At high-temperature, $|{\rm TF}_3(\beta)\ra$ is equivalent to an equal superposition of all possible product states in which each copy occupies the same quantum state
\begin{equation}
\label{rho3ghz}
|{\rm TF}_3(\beta=0)\ra=\frac{1}{d^{\frac{1}{2}}}\sum_{k}|k\rangle|k\rangle|k\rangle,
\end{equation}%
and in this sense Eq. (\ref{rho3ghz}) is a generalization of the GHZ state.
Note that in the standard GHZ state of three qubits $|{\rm GHZ}\ra=[|000\ra+|111\ra]/\sqrt{2}$, tracing over any of the subsystems gives rise to the maximally mixed state
\beqa
\rho=\frac{1}{2}(|00\ra\la 00|+|11\ra\la11|),
\eeqa
analogous to $\rho(\infty)$ in Eq. (\ref{rho2inf}). This is an unentangled mixed state including only classical correlations between the two copies.

We can consider the dynamics of the  ``thermofield triple state'' (\ref{tf3}) under decoherence that would give rise to a new fix point of evolution
\begin{equation}
\rho _{3}(\infty)=\frac{1}{Z(\beta )}\sum_{k}e^{-\beta E_{k}}|k\rangle
|k\rangle|k\rangle  \langle k|\langle k|\langle k|, 
\end{equation}%
that can be as well purified as
\begin{equation}
\label{tf4}
|{\rm TF}_4(\beta)\ra=\frac{1}{[Z(\beta )]^{\frac{1}{2}}}\sum_{k}e^{-\frac{\beta}{2} E_{k}} |k\rangle^{\otimes 4}.
\end{equation}%

Clearly, a thermofield $n$-tuple state involving $n$ copies of a system
\begin{equation}
\label{tf4n}
|{\rm TF} _n(\beta)\ra=\frac{1}{[Z(\beta )]^{\frac{1}{2}}}\sum_{k}e^{-\frac{\beta}{2} E_{k}} |k\rangle^{\otimes n}.
\end{equation}%
decoheres into a mixed state
\begin{equation}
\rho _{n}(\infty)=\frac{1}{Z(\beta )}\sum_{k}e^{-\beta E_{k}}
(|k\rangle\langle k|)^{\otimes n},
\end{equation}%
whose purification is of the form $|{\rm TF}_{n+1}(\beta)\ra$.
Note that the purity of 
\beqa
\tr[\rho _{n}(\infty)^2]=\frac{Z(2\beta)}{Z(\beta)^2}=\tr[\rho_{2}(\infty)^2].
\eeqa
It has been known that a holographic dual of a GHZ state cannot be describes by a classical gravity solution
\cite{Susskind16}. Our  result above suggests that a possible gravity dual involves random noises or equally random boundary conditions.

Moreover, as we mentioned, we can purify the whole mixed time-evolution (\ref{decKra}) under decoherence by adding the third CFT Hilbert space as a unitary time-evolution. 
In this case, at $t=0$, since the original two CFTs are in the pure state described by the thermo-field double state, we can choose the third CFT to be simply in the vacuum state. After the time evolution,
the decoherent fluctuations induce the quantum entanglement between the two CFTs and the third CFT, and the global state finally approaches the GHZ-like state (\ref{tf3}).

\begin{figure}[t]
  \centering
  \includegraphics[width=3.15in]{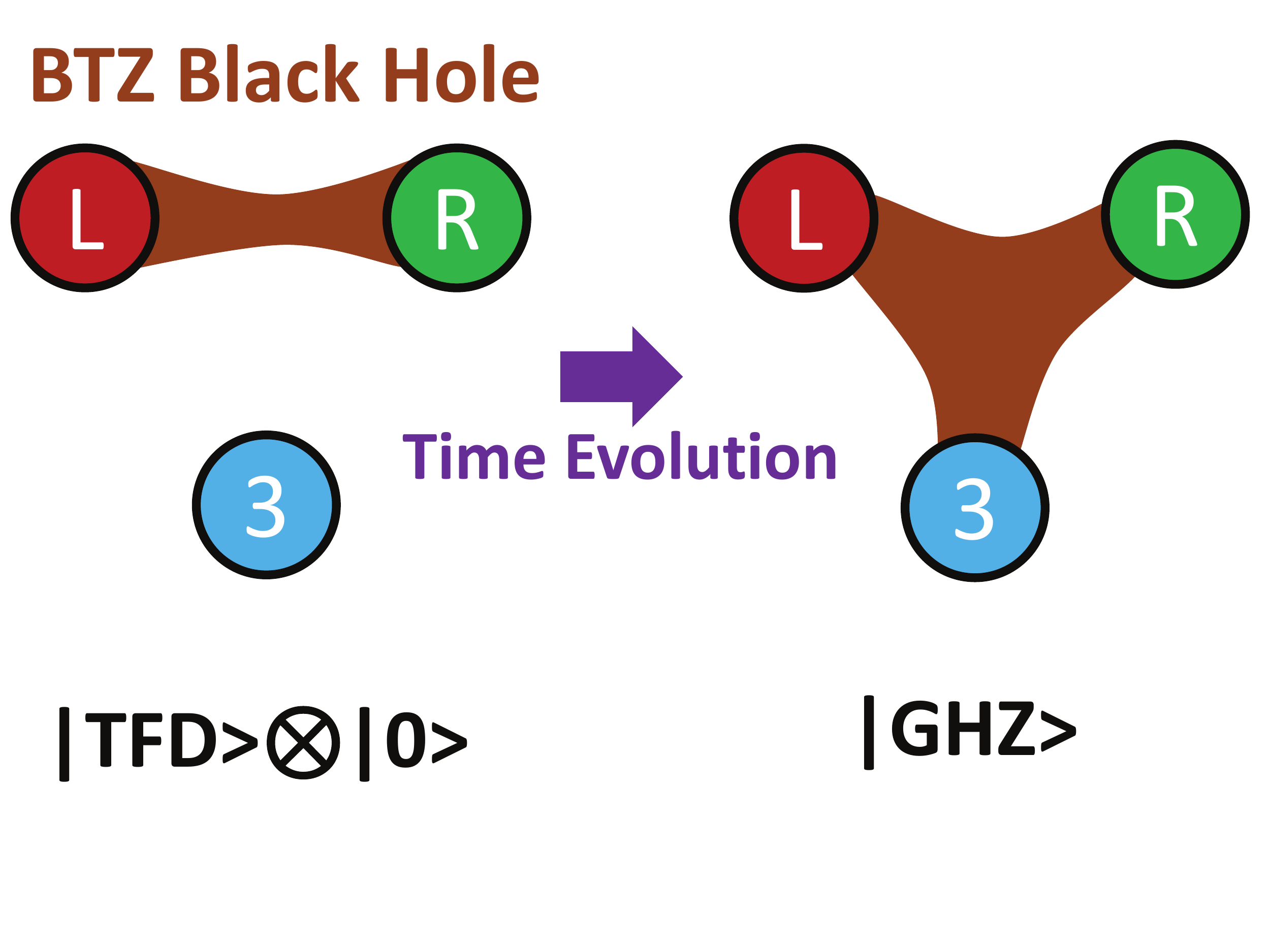}
  \caption{A sketch of time evolution from a thermofield state into GHZ state in AdS/CFT.}
\label{fig:ghz}
\end{figure}

This evolution is interpreted from the viewpoint of a gravity dual as follows.
We start with an eternal BTZ black hole and a single pure AdS at $t=0$. Then we introduce interactions between 
these two spacetimes, which are dual to the decoherent fluctuations.  This leads to a time evolution of the geometry with the three asymptotic AdS boundaries. In this process, the originally disconnected two universes, an eternal BTZ and a pure AdS, are glued with each other. 
Finally they evolves into a gravity dual of the GHZ-like state. This is sketched in Fig. \ref{fig:ghz}.
However, notice that since a classical geometry cannot describe a GHZ-like state \cite{Susskind16}, 
we expect the final spacetime $t=\infty$ gets highly quantum.

\section{Decoherence in Boundary States}
So far we have analyzed the role of decoherence  by focusing on two entangled copies of the CFT, i.e., the thermofield double. As a setting to explore decoherence for a pure state in a single 2d CFT, we next consider the time evolution of a boundary state $|B\ra$, so-called Cardy state \cite{Cardy89}. The dynamics of the latter has so far been studied in the  context of quantum quenches under unitary dynamics \cite{Cardy14,Cardy16}. In the AdS/CFT, this setup corresponds to the time evolution of single sided black hole in AdS \cite{HaMa13}. It is an intriguing prospect to ask whether one can find a decoherence  process which 
is analogous to the previous one we found for the thermofield double state dual to the two sided eternal black hole.

We can separate the CFT Hamiltonian into the left (chiral)  and right-moving (anti-chiral) part, $H=H_l+H_r$, and introduce  decoherence by considering the fluctuating operators 
$\delta H_l=\s{\gamma}\xi_l H_l$ and $\delta H_r=\s{\gamma}\xi_r H_r$, where $\xi_{l}$ and $\xi_{r}$ 
are independent Gaussian noises.  

In general, the Cardy states $|B_a\ra$ labeled by the index $a$ is a linear combination of Ishibashi states $|I_\alpha\ra$ \cite{Ishibashi89}, which are labeled by the index $|\alpha\ra$ of primary states.   The Ishibashi state $|I_\alpha\ra$ is given by an infinite summation of the descendants of the primary state $|\alpha\ra$, which has the conformal dimension $h_\alpha(=\bar{h}_\alpha)$. A Cardy state 
is expressed as 
\ba
|B_a\ra=\sum_\alpha c^\alpha_a |I_\alpha\ra,
\ea
where $c^\alpha_a$ are complex coefficients.

Since the Ishibashi state is  the maximally entanglement state of the descendants
\ba
|I_\alpha\ra=\sum_{p\in \mbox{descendants}}|p,\alpha\ra_L |p,\alpha\ra_R,
\ea 
the norm $\la I_\alpha|I_\alpha\ra$ is 
divergent. Accordingly, we need a regularization for the Cardy state and we can make this by setting 
\ba
|\Psi_a\ra=\frac{1}{\s{Z_{bry}(\beta)}}e^{-\frac{\beta}{4}H}|B_a\ra,
\ea 
where 
\ba
Z_{bry}(\beta)=\la B_a|e^{-\frac{\beta}{2}H}|B_a\ra,
\ea
which is the partition function on the cylinder, this is, the open string thermal partition function.

We consider a time evolution which starts from this state $|\Psi_a\ra$ at $t=0$ and take into account the mentioned decoherence. The time evolution of density matrix in this CFT is described by
\ba
\rho(t)=\frac{1}{Z_{bry}(\beta)}\sum_{\alpha,\beta}\sum_{p,q} c^{\alpha}_a c^{*\beta}_b e^{-\frac{\beta}{2}(E^l_{p,\alpha}+E^l_{q,\beta})}
 e^{-2it(E^l_{p,\alpha}-E^l_{q,\beta})} e^{-\gamma t(E^l_{p,\alpha}-E^l_{q,\beta})^2},
\ea
where $E^l_{p,\alpha}$ denotes the eigenvalue of $H_l$ of the state $|p,\alpha\lb$.

For example, the purity is expressed as follows
\ba
\mbox{Tr}[\rho(t)^2]=\frac{1}{Z_{bry}(\beta)^2}\sum_{\alpha,\beta}\sum_{p,q}|c^{\alpha}_a|^2|c^{\beta}_b|^2 
 e^{-\beta(E^l_{p,\alpha}+E^l_{q,\beta})} e^{-2\gamma t(E^l_{p,\alpha}-E^l_{q,\beta})^2}.
\ea

Thus, when there is no degeneracy in the energy spectrum, we obtain in late time limit $t\to \infty$ 
\ba
\mbox{Tr}[\rho(\infty)^2]=\frac{1}{Z_{bry}(\beta)^2}\sum_{\alpha}\sum_{p}|c^{\alpha}_a|^4 
 e^{-2\beta E^l_{p,\alpha}}.
\ea

In the same way, in the absence of degeneracy, in the late time limit we find 
\ba
\mbox{Tr}[\rho(\infty)^n]=\frac{1}{Z_{bry}(\beta)^n}\sum_{\alpha}\sum_{p}|c^{\alpha}_a|^{2n} 
 e^{-n\beta E^l_{p,\alpha}}.
\ea 

Then, the von Neumann entropy reads 
\ba
S[\rho(\infty)]&=&-\mbox{Tr}[\rho(\infty)\log\rho(\infty)]\no
&=& \log Z_{bry}(\beta)+\beta \la E_l\ra_{\beta}-\frac{\sum_{\alpha,p}|c^\alpha_a|^2 \log|c^\alpha_a|^2\cdot  e^{-\beta E^l_{p,\alpha}}}{\sum_{\alpha,p}|c^\alpha_a|^2 e^{-\beta E^l_{p,\alpha}}}\no
&=& S_{th,bdy}-\frac{\sum_{\alpha,p}|c^\alpha_a|^2 \log|c^\alpha_a|^2  e^{-\beta E^L_{p,\alpha}}}{\sum_{\alpha,p}|c^\alpha_a|^2 e^{-\beta E^l_{p,\alpha}}}, \label{srhoz}
\ea
where $S_{th,bdy}$ is the thermal entropy of the boundary conformal field theory (BCFT).
Note that at $t=0$ the state is pure (i.e., given by $|\Psi_a\ra$) and thus $S[\rho(0)]=0$. Therefore the above entropy 
$S[\rho(\infty)]$ is the entropy which corresponds to the entanglement between the CFT and the thermal bath, 
produced by the decoherence.

Now let us consider the quantum entanglement between the left and right-moving sector \cite{DasDatta15}.
The reduced density matrix for the left-moving sector is computed as 
\ba
\rho_l(t)=\mbox{Tr}_r[\rho(t)]=\frac{1}{Z_{bdy}(\beta)}\sum_{\alpha,p}|c^\alpha_a|^2e^{-\beta E^l_{p,\alpha}}|p,\alpha\ra\la p,\alpha|,
\ea 
which is time-independent. This leads to the entanglement entropy between the left and right-moving sector given by 
\ba
S_l= S_{th,bdy}-\frac{\sum_{\alpha,p}|c^\alpha_a|^2 \log|c^\alpha_a|^2  e^{-\beta E^l_{p,\alpha}}}{\sum_{\alpha,p}|c^\alpha_a|^2 e^{-\beta E^l_{p,\alpha}}},
\ea
which coincides with $S[\rho(\infty)]$ given by (\ref{srhoz}).  This is very natural as the original quantum entanglement between left-right modes turned into the classical correlations due to the entanglement with the thermal bath.

 \section{Summary}
 
 We have analyzed the decoherence dynamics induced by energy dephasing in conformal field theories. To this end, we have first shown that the time evolution of the purity  and R\'enyi entropies of an initial thermofield double state is characterized by a monotonic decay from unity to a thermal asymptotic value. This time dependence  is as well shared by the logarithmic negativity, whose long-time asymptotics crucially depends on the presence of degeneracies in the energy spectrum. 
In their absence, the asymptotic logarithmic negative can be shown to vanish and the late-time state  involves only classical correlations between the two CFT copies. By contrast, degeneracies lead to a non-zero quantum entanglement between the  two copies that survives at long times  in the  decohered state, as shown explicitly  using two-dimensional Dirac fermion CFTs as a test-bed.
Aiming at the  characterization of the gravity dual associated with a TFD under decoherence, an analysis of the two-point functions suggests the shrinking of the inner horizon regions of the eternal BTZ. 
We have further analyzed the information loss induced by decoherence as a result of its leakage to a reference environment and proposed a gravitational dual interpretation of the decoherent dynamics.
Via the AdS/CFT, this setup with the environment implies how a gravity dual of GHZ state looks like.
Finally, we have shown the effect of decoherence on a single CFT copy by analyzing the dynamics of boundary states.

\section*{Acknowledgements} We are grateful to Herman Verlinde for useful correspondence.
TT is supported by the Simons Foundation through the ``It from Qubit'' collaboration. TT is supported by JSPS Grant-in-Aid for Scientific Research (A) No.16H02182 and by JSPS Grant-in-Aid for Challenging Research (Exploratory) 18K18766. TT is also supported by World Premier International Research Center Initiative (WPI Initiative) from the Japan Ministry of Education, Culture, Sports, Science and Technology (MEXT). We are grateful to the long term workshop Quantum Information and String Theory (YITP-T-19-03) held at Yukawa Institute for Theoretical Physics, Kyoto University, where this work was initiated.

\bibliography{DecoCFTlib.bib}	

\begin{thebibliography}{44}
\expandafter\ifx\csname natexlab\endcsname\relax\def\natexlab#1{#1}\fi
\expandafter\ifx\csname bibnamefont\endcsname\relax
  \def\bibnamefont#1{#1}\fi
\expandafter\ifx\csname bibfnamefont\endcsname\relax
  \def\bibfnamefont#1{#1}\fi
\expandafter\ifx\csname citenamefont\endcsname\relax
  \def\citenamefont#1{#1}\fi
\expandafter\ifx\csname url\endcsname\relax
  \def\url#1{\texttt{#1}}\fi
\expandafter\ifx\csname urlprefix\endcsname\relax\def\urlprefix{URL }\fi
\providecommand{\bibinfo}[2]{#2}
\providecommand{\eprint}[2][]{\url{#2}}

\bibitem[{\citenamefont{Zurek}(2003)}]{Zurek03}
\bibinfo{author}{\bibfnamefont{W.~H.} \bibnamefont{Zurek}},
  \bibinfo{journal}{Rev. Mod. Phys.} \textbf{\bibinfo{volume}{75}},
  \bibinfo{pages}{715} (\bibinfo{year}{2003}),
  \urlprefix\url{https://link.aps.org/doi/10.1103/RevModPhys.75.715}.

\bibitem[{\citenamefont{Egusquiza et~al.}(1999)\citenamefont{Egusquiza, Garay,
  and Raya}}]{Egusquiza99}
\bibinfo{author}{\bibfnamefont{I.~L.} \bibnamefont{Egusquiza}},
  \bibinfo{author}{\bibfnamefont{L.~J.} \bibnamefont{Garay}}, \bibnamefont{and}
  \bibinfo{author}{\bibfnamefont{J.~M.} \bibnamefont{Raya}},
  \bibinfo{journal}{Phys. Rev. A} \textbf{\bibinfo{volume}{59}},
  \bibinfo{pages}{3236} (\bibinfo{year}{1999}),
  \urlprefix\url{https://link.aps.org/doi/10.1103/PhysRevA.59.3236}.

\bibitem[{\citenamefont{Gambini et~al.}(2007)\citenamefont{Gambini, Porto, and
  Pullin}}]{Gambini07}
\bibinfo{author}{\bibfnamefont{R.}~\bibnamefont{Gambini}},
  \bibinfo{author}{\bibfnamefont{R.~A.} \bibnamefont{Porto}}, \bibnamefont{and}
  \bibinfo{author}{\bibfnamefont{J.}~\bibnamefont{Pullin}},
  \bibinfo{journal}{General Relativity and Gravitation}
  \textbf{\bibinfo{volume}{39}}, \bibinfo{pages}{1143} (\bibinfo{year}{2007}),
  ISSN \bibinfo{issn}{1572-9532},
  \urlprefix\url{https://doi.org/10.1007/s10714-007-0451-1}.

\bibitem[{\citenamefont{Gisin}(1984)}]{Gisin84}
\bibinfo{author}{\bibfnamefont{N.}~\bibnamefont{Gisin}},
  \bibinfo{journal}{Phys. Rev. Lett.} \textbf{\bibinfo{volume}{52}},
  \bibinfo{pages}{1657} (\bibinfo{year}{1984}),
  \urlprefix\url{https://link.aps.org/doi/10.1103/PhysRevLett.52.1657}.

\bibitem[{\citenamefont{Percival}(1994)}]{Percival94}
\bibinfo{author}{\bibfnamefont{I.~C.} \bibnamefont{Percival}},
  \bibinfo{journal}{Proceedings of the Royal Society of London. Series A:
  Mathematical and Physical Sciences} \textbf{\bibinfo{volume}{447}},
  \bibinfo{pages}{189} (\bibinfo{year}{1994}),
  \urlprefix\url{https://royalsocietypublishing.org/doi/abs/10.1098/rspa.1994.0135}.

\bibitem[{\citenamefont{Adler}(2003)}]{Adler03}
\bibinfo{author}{\bibfnamefont{S.~L.} \bibnamefont{Adler}},
  \bibinfo{journal}{Phys. Rev. D} \textbf{\bibinfo{volume}{67}},
  \bibinfo{pages}{025007} (\bibinfo{year}{2003}),
  \urlprefix\url{https://link.aps.org/doi/10.1103/PhysRevD.67.025007}.

\bibitem[{\citenamefont{Bassi and Ghirardi}(2003)}]{Bassi03}
\bibinfo{author}{\bibfnamefont{A.}~\bibnamefont{Bassi}} \bibnamefont{and}
  \bibinfo{author}{\bibfnamefont{G.}~\bibnamefont{Ghirardi}},
  \bibinfo{journal}{Physics Reports} \textbf{\bibinfo{volume}{379}},
  \bibinfo{pages}{257 } (\bibinfo{year}{2003}), ISSN \bibinfo{issn}{0370-1573},
  \urlprefix\url{http://www.sciencedirect.com/science/article/pii/S0370157303001030}.

\bibitem[{\citenamefont{Bassi et~al.}(2013)\citenamefont{Bassi, Lochan, Satin,
  Singh, and Ulbricht}}]{Bassi13}
\bibinfo{author}{\bibfnamefont{A.}~\bibnamefont{Bassi}},
  \bibinfo{author}{\bibfnamefont{K.}~\bibnamefont{Lochan}},
  \bibinfo{author}{\bibfnamefont{S.}~\bibnamefont{Satin}},
  \bibinfo{author}{\bibfnamefont{T.~P.} \bibnamefont{Singh}}, \bibnamefont{and}
  \bibinfo{author}{\bibfnamefont{H.}~\bibnamefont{Ulbricht}},
  \bibinfo{journal}{Rev. Mod. Phys.} \textbf{\bibinfo{volume}{85}},
  \bibinfo{pages}{471} (\bibinfo{year}{2013}),
  \urlprefix\url{https://link.aps.org/doi/10.1103/RevModPhys.85.471}.

\bibitem[{\citenamefont{Milburn}(1991)}]{Milburn91}
\bibinfo{author}{\bibfnamefont{G.~J.} \bibnamefont{Milburn}},
  \bibinfo{journal}{Phys. Rev. A} \textbf{\bibinfo{volume}{44}},
  \bibinfo{pages}{5401} (\bibinfo{year}{1991}),
  \urlprefix\url{https://link.aps.org/doi/10.1103/PhysRevA.44.5401}.

\bibitem[{\citenamefont{Chenu et~al.}(2017)\citenamefont{Chenu, Beau, Cao, and
  del Campo}}]{Chenu17}
\bibinfo{author}{\bibfnamefont{A.}~\bibnamefont{Chenu}},
  \bibinfo{author}{\bibfnamefont{M.}~\bibnamefont{Beau}},
  \bibinfo{author}{\bibfnamefont{J.}~\bibnamefont{Cao}}, \bibnamefont{and}
  \bibinfo{author}{\bibfnamefont{A.}~\bibnamefont{del Campo}},
  \bibinfo{journal}{Phys. Rev. Lett.} \textbf{\bibinfo{volume}{118}},
  \bibinfo{pages}{140403} (\bibinfo{year}{2017}),
  \urlprefix\url{https://link.aps.org/doi/10.1103/PhysRevLett.118.140403}.

\bibitem[{\citenamefont{Korbicz et~al.}(2017)\citenamefont{Korbicz, Aguilar,
  \ifmmode \acute{C}\else \'{C}\fi{}wikli\ifmmode~\acute{n}\else \'{n}\fi{}ski,
  and Horodecki}}]{Korbicz17}
\bibinfo{author}{\bibfnamefont{J.~K.} \bibnamefont{Korbicz}},
  \bibinfo{author}{\bibfnamefont{E.~A.} \bibnamefont{Aguilar}},
  \bibinfo{author}{\bibfnamefont{P.}~\bibnamefont{\ifmmode \acute{C}\else
  \'{C}\fi{}wikli\ifmmode~\acute{n}\else \'{n}\fi{}ski}}, \bibnamefont{and}
  \bibinfo{author}{\bibfnamefont{P.}~\bibnamefont{Horodecki}},
  \bibinfo{journal}{Phys. Rev. A} \textbf{\bibinfo{volume}{96}},
  \bibinfo{pages}{032124} (\bibinfo{year}{2017}),
  \urlprefix\url{https://link.aps.org/doi/10.1103/PhysRevA.96.032124}.

\bibitem[{\citenamefont{Maldacena}(1999)}]{Maldacena99}
\bibinfo{author}{\bibfnamefont{J.}~\bibnamefont{Maldacena}},
  \bibinfo{journal}{International Journal of Theoretical Physics}
  \textbf{\bibinfo{volume}{38}}, \bibinfo{pages}{1113} (\bibinfo{year}{1999}),
  ISSN \bibinfo{issn}{1572-9575},
  \urlprefix\url{https://doi.org/10.1023/A:1026654312961}.

\bibitem[{\citenamefont{Ryu and Takayanagi}(2006{\natexlab{a}})}]{RT06}
\bibinfo{author}{\bibfnamefont{S.}~\bibnamefont{Ryu}} \bibnamefont{and}
  \bibinfo{author}{\bibfnamefont{T.}~\bibnamefont{Takayanagi}},
  \bibinfo{journal}{Phys. Rev. Lett.} \textbf{\bibinfo{volume}{96}},
  \bibinfo{pages}{181602} (\bibinfo{year}{2006}{\natexlab{a}}),
  \urlprefix\url{https://link.aps.org/doi/10.1103/PhysRevLett.96.181602}.

\bibitem[{\citenamefont{Ryu and Takayanagi}(2006{\natexlab{b}})}]{RT06b}
\bibinfo{author}{\bibfnamefont{S.}~\bibnamefont{Ryu}} \bibnamefont{and}
  \bibinfo{author}{\bibfnamefont{T.}~\bibnamefont{Takayanagi}},
  \bibinfo{journal}{Journal of High Energy Physics}
  \textbf{\bibinfo{volume}{2006}}, \bibinfo{pages}{045}
  (\bibinfo{year}{2006}{\natexlab{b}}),
  \urlprefix\url{https://doi.org/10.1088%2F1126-6708%2F2006%2F08%2F045}.

\bibitem[{\citenamefont{Hubeny et~al.}(2007)\citenamefont{Hubeny, Rangamani,
  and Takayanagi}}]{Hubeny07}
\bibinfo{author}{\bibfnamefont{V.~E.} \bibnamefont{Hubeny}},
  \bibinfo{author}{\bibfnamefont{M.}~\bibnamefont{Rangamani}},
  \bibnamefont{and}
  \bibinfo{author}{\bibfnamefont{T.}~\bibnamefont{Takayanagi}},
  \bibinfo{journal}{Journal of High Energy Physics}
  \textbf{\bibinfo{volume}{2007}}, \bibinfo{pages}{062} (\bibinfo{year}{2007}),
  \urlprefix\url{https://doi.org/10.1088%2F1126-6708%2F2007%2F07%2F062}.

\bibitem[{\citenamefont{Umemoto and Takayanagi}(2018)}]{Umemoto18}
\bibinfo{author}{\bibfnamefont{K.}~\bibnamefont{Umemoto}} \bibnamefont{and}
  \bibinfo{author}{\bibfnamefont{T.}~\bibnamefont{Takayanagi}},
  \bibinfo{journal}{Nature Physics} \textbf{\bibinfo{volume}{14}},
  \bibinfo{pages}{573} (\bibinfo{year}{2018}),
  \urlprefix\url{https://doi.org/10.1038/s41567-018-0075-2}.

\bibitem[{\citenamefont{Nguyen et~al.}(2018)\citenamefont{Nguyen, Devakul,
  Halbasch, Zaletel, and Swingle}}]{Nguyen18}
\bibinfo{author}{\bibfnamefont{P.}~\bibnamefont{Nguyen}},
  \bibinfo{author}{\bibfnamefont{T.}~\bibnamefont{Devakul}},
  \bibinfo{author}{\bibfnamefont{M.~G.} \bibnamefont{Halbasch}},
  \bibinfo{author}{\bibfnamefont{M.~P.} \bibnamefont{Zaletel}},
  \bibnamefont{and} \bibinfo{author}{\bibfnamefont{B.}~\bibnamefont{Swingle}},
  \bibinfo{journal}{Journal of High Energy Physics}
  \textbf{\bibinfo{volume}{2018}}, \bibinfo{pages}{98} (\bibinfo{year}{2018}),
  ISSN \bibinfo{issn}{1029-8479},
  \urlprefix\url{https://doi.org/10.1007/JHEP01(2018)098}.

\bibitem[{\citenamefont{Lindblad}(1976)}]{Lindblad76}
\bibinfo{author}{\bibfnamefont{G.}~\bibnamefont{Lindblad}},
  \bibinfo{journal}{Communications in Mathematical Physics}
  \textbf{\bibinfo{volume}{48}}, \bibinfo{pages}{119} (\bibinfo{year}{1976}),
  ISSN \bibinfo{issn}{1432-0916},
  \urlprefix\url{https://doi.org/10.1007/BF01608499}.

\bibitem[{\citenamefont{Beau et~al.}(2017)\citenamefont{Beau, Kiukas,
  Egusquiza, and del Campo}}]{Beau17}
\bibinfo{author}{\bibfnamefont{M.}~\bibnamefont{Beau}},
  \bibinfo{author}{\bibfnamefont{J.}~\bibnamefont{Kiukas}},
  \bibinfo{author}{\bibfnamefont{I.~L.} \bibnamefont{Egusquiza}},
  \bibnamefont{and} \bibinfo{author}{\bibfnamefont{A.}~\bibnamefont{del
  Campo}}, \bibinfo{journal}{Phys. Rev. Lett.} \textbf{\bibinfo{volume}{119}},
  \bibinfo{pages}{130401} (\bibinfo{year}{2017}),
  \urlprefix\url{https://link.aps.org/doi/10.1103/PhysRevLett.119.130401}.

\bibitem[{\citenamefont{Xu et~al.}(2019)\citenamefont{Xu, Garc\'{\i}a-Pintos,
  Chenu, and del Campo}}]{Xu19}
\bibinfo{author}{\bibfnamefont{Z.}~\bibnamefont{Xu}},
  \bibinfo{author}{\bibfnamefont{L.~P.} \bibnamefont{Garc\'{\i}a-Pintos}},
  \bibinfo{author}{\bibfnamefont{A.}~\bibnamefont{Chenu}}, \bibnamefont{and}
  \bibinfo{author}{\bibfnamefont{A.}~\bibnamefont{del Campo}},
  \bibinfo{journal}{Phys. Rev. Lett.} \textbf{\bibinfo{volume}{122}},
  \bibinfo{pages}{014103} (\bibinfo{year}{2019}),
  \urlprefix\url{https://link.aps.org/doi/10.1103/PhysRevLett.122.014103}.

\bibitem[{\citenamefont{Umezawa et~al.}(1982)\citenamefont{Umezawa, Matsumoto,
  and Tachiki}}]{Umezawa82}
\bibinfo{author}{\bibfnamefont{H.}~\bibnamefont{Umezawa}},
  \bibinfo{author}{\bibfnamefont{H.}~\bibnamefont{Matsumoto}},
  \bibnamefont{and} \bibinfo{author}{\bibfnamefont{M.}~\bibnamefont{Tachiki}},
  \emph{\bibinfo{title}{ThermoField Dynamics and Condensed States}}
  (\bibinfo{publisher}{North-Holland}, \bibinfo{year}{1982}).

\bibitem[{\citenamefont{Maldacena and Susskind}(2013)}]{Maldacena13}
\bibinfo{author}{\bibfnamefont{J.}~\bibnamefont{Maldacena}} \bibnamefont{and}
  \bibinfo{author}{\bibfnamefont{L.}~\bibnamefont{Susskind}},
  \bibinfo{journal}{Fortschritte der Physik} \textbf{\bibinfo{volume}{61}},
  \bibinfo{pages}{781} (\bibinfo{year}{2013}),
  \urlprefix\url{https://onlinelibrary.wiley.com/doi/abs/10.1002/prop.201300020}.

\bibitem[{\citenamefont{Shenker and Stanford}(2014)}]{Shenker14}
\bibinfo{author}{\bibfnamefont{S.~H.} \bibnamefont{Shenker}} \bibnamefont{and}
  \bibinfo{author}{\bibfnamefont{D.}~\bibnamefont{Stanford}},
  \bibinfo{journal}{Journal of High Energy Physics}
  \textbf{\bibinfo{volume}{2014}}, \bibinfo{pages}{67} (\bibinfo{year}{2014}),
  ISSN \bibinfo{issn}{1029-8479},
  \urlprefix\url{https://doi.org/10.1007/JHEP03(2014)067}.

\bibitem[{\citenamefont{Pastawski et~al.}(2017)\citenamefont{Pastawski, Eisert,
  and Wilming}}]{Pastawski17}
\bibinfo{author}{\bibfnamefont{F.}~\bibnamefont{Pastawski}},
  \bibinfo{author}{\bibfnamefont{J.}~\bibnamefont{Eisert}}, \bibnamefont{and}
  \bibinfo{author}{\bibfnamefont{H.}~\bibnamefont{Wilming}},
  \bibinfo{journal}{Phys. Rev. Lett.} \textbf{\bibinfo{volume}{119}},
  \bibinfo{pages}{020501} (\bibinfo{year}{2017}),
  \urlprefix\url{https://link.aps.org/doi/10.1103/PhysRevLett.119.020501}.

\bibitem[{\citenamefont{Nagasawa}(2000)}]{Nagasawa00}
\bibinfo{author}{\bibfnamefont{M.}~\bibnamefont{Nagasawa}},
  \emph{\bibinfo{title}{Stochastic Processes in Quantum Physics}}
  (\bibinfo{publisher}{Birkhauser, Boston}, \bibinfo{year}{2000}).

\bibitem[{\citenamefont{Lidar et~al.}(2006)\citenamefont{Lidar, Shabani, and
  Alicki}}]{Lidar06}
\bibinfo{author}{\bibfnamefont{D.}~\bibnamefont{Lidar}},
  \bibinfo{author}{\bibfnamefont{A.}~\bibnamefont{Shabani}}, \bibnamefont{and}
  \bibinfo{author}{\bibfnamefont{R.}~\bibnamefont{Alicki}},
  \bibinfo{journal}{Chemical Physics} \textbf{\bibinfo{volume}{322}},
  \bibinfo{pages}{82 } (\bibinfo{year}{2006}), ISSN \bibinfo{issn}{0301-0104},
  \bibinfo{note}{real-time dynamics in complex quantum systems},
  \urlprefix\url{http://www.sciencedirect.com/science/article/pii/S0301010405002752}.

\bibitem[{\citenamefont{Cotler et~al.}(2017)\citenamefont{Cotler, Gur-Ari,
  Hanada, Polchinski, Saad, Shenker, Stanford, Streicher, and
  Tezuka}}]{Cotler17}
\bibinfo{author}{\bibfnamefont{J.~S.} \bibnamefont{Cotler}},
  \bibinfo{author}{\bibfnamefont{G.}~\bibnamefont{Gur-Ari}},
  \bibinfo{author}{\bibfnamefont{M.}~\bibnamefont{Hanada}},
  \bibinfo{author}{\bibfnamefont{J.}~\bibnamefont{Polchinski}},
  \bibinfo{author}{\bibfnamefont{P.}~\bibnamefont{Saad}},
  \bibinfo{author}{\bibfnamefont{S.~H.} \bibnamefont{Shenker}},
  \bibinfo{author}{\bibfnamefont{D.}~\bibnamefont{Stanford}},
  \bibinfo{author}{\bibfnamefont{A.}~\bibnamefont{Streicher}},
  \bibnamefont{and} \bibinfo{author}{\bibfnamefont{M.}~\bibnamefont{Tezuka}},
  \bibinfo{journal}{Journal of High Energy Physics}
  \textbf{\bibinfo{volume}{2017}}, \bibinfo{pages}{118} (\bibinfo{year}{2017}),
  ISSN \bibinfo{issn}{1029-8479},
  \urlprefix\url{https://doi.org/10.1007/JHEP05(2017)118}.

\bibitem[{\citenamefont{Dyer and Gur-Ari}(2017)}]{Dyer17}
\bibinfo{author}{\bibfnamefont{E.}~\bibnamefont{Dyer}} \bibnamefont{and}
  \bibinfo{author}{\bibfnamefont{G.}~\bibnamefont{Gur-Ari}},
  \bibinfo{journal}{Journal of High Energy Physics}
  \textbf{\bibinfo{volume}{2017}}, \bibinfo{pages}{75} (\bibinfo{year}{2017}),
  ISSN \bibinfo{issn}{1029-8479},
  \urlprefix\url{https://doi.org/10.1007/JHEP08(2017)075}.

\bibitem[{\citenamefont{del Campo et~al.}(2017)\citenamefont{del Campo,
  Molina-Vilaplana, and Sonner}}]{delcampo17}
\bibinfo{author}{\bibfnamefont{A.}~\bibnamefont{del Campo}},
  \bibinfo{author}{\bibfnamefont{J.}~\bibnamefont{Molina-Vilaplana}},
  \bibnamefont{and} \bibinfo{author}{\bibfnamefont{J.}~\bibnamefont{Sonner}},
  \bibinfo{journal}{Phys. Rev. D} \textbf{\bibinfo{volume}{95}},
  \bibinfo{pages}{126008} (\bibinfo{year}{2017}),
  \urlprefix\url{https://link.aps.org/doi/10.1103/PhysRevD.95.126008}.

\bibitem[{\citenamefont{Plenio}(2005)}]{Plenio05}
\bibinfo{author}{\bibfnamefont{M.~B.} \bibnamefont{Plenio}},
  \bibinfo{journal}{Phys. Rev. Lett.} \textbf{\bibinfo{volume}{95}},
  \bibinfo{pages}{090503} (\bibinfo{year}{2005}),
  \urlprefix\url{https://link.aps.org/doi/10.1103/PhysRevLett.95.090503}.

\bibitem[{\citenamefont{Calabrese et~al.}(2012)\citenamefont{Calabrese, Cardy,
  and Tonni}}]{CCT12}
\bibinfo{author}{\bibfnamefont{P.}~\bibnamefont{Calabrese}},
  \bibinfo{author}{\bibfnamefont{J.}~\bibnamefont{Cardy}}, \bibnamefont{and}
  \bibinfo{author}{\bibfnamefont{E.}~\bibnamefont{Tonni}},
  \bibinfo{journal}{Phys. Rev. Lett.} \textbf{\bibinfo{volume}{109}},
  \bibinfo{pages}{130502} (\bibinfo{year}{2012}),
  \urlprefix\url{https://link.aps.org/doi/10.1103/PhysRevLett.109.130502}.

\bibitem[{\citenamefont{Azeyanagi et~al.}(2008)\citenamefont{Azeyanagi,
  Nishioka, and Takayanagi}}]{Azeyanagi08}
\bibinfo{author}{\bibfnamefont{T.}~\bibnamefont{Azeyanagi}},
  \bibinfo{author}{\bibfnamefont{T.}~\bibnamefont{Nishioka}}, \bibnamefont{and}
  \bibinfo{author}{\bibfnamefont{T.}~\bibnamefont{Takayanagi}},
  \bibinfo{journal}{Phys. Rev. D} \textbf{\bibinfo{volume}{77}},
  \bibinfo{pages}{064005} (\bibinfo{year}{2008}),
  \urlprefix\url{https://link.aps.org/doi/10.1103/PhysRevD.77.064005}.

\bibitem[{\citenamefont{Maldacena and Strominger}(1998)}]{Strominger98}
\bibinfo{author}{\bibfnamefont{J.}~\bibnamefont{Maldacena}} \bibnamefont{and}
  \bibinfo{author}{\bibfnamefont{A.}~\bibnamefont{Strominger}},
  \bibinfo{journal}{Journal of High Energy Physics}
  \textbf{\bibinfo{volume}{1998}}, \bibinfo{pages}{005} (\bibinfo{year}{1998}),
  \urlprefix\url{https://doi.org/10.1088%2F1126-6708%2F1998%2F12%2F005}.

\bibitem[{\citenamefont{Hayashi}(2010)}]{Hayashi10}
\bibinfo{author}{\bibfnamefont{M.}~\bibnamefont{Hayashi}},
  \emph{\bibinfo{title}{Quantum Information}} (\bibinfo{publisher}{Springer,
  Heidelberg}, \bibinfo{year}{2010}).

\bibitem[{\citenamefont{Kraus}(1971)}]{Kraus71}
\bibinfo{author}{\bibfnamefont{K.}~\bibnamefont{Kraus}},
  \bibinfo{journal}{Annals of Physics} \textbf{\bibinfo{volume}{64}},
  \bibinfo{pages}{311 } (\bibinfo{year}{1971}), ISSN \bibinfo{issn}{0003-4916},
  \urlprefix\url{http://www.sciencedirect.com/science/article/pii/0003491671901084}.

\bibitem[{\citenamefont{Verlinde}(2019)}]{Verlinde19}
\bibinfo{author}{\bibfnamefont{H.}~\bibnamefont{Verlinde}}
  (\bibinfo{year}{2019}),
  \urlprefix\url{http://www2.yukawa.kyoto-u.ac.jp/~qist2019/slides/5th/Verlinde.pdf}.

\bibitem[{\citenamefont{Almheiri et~al.}(2019)\citenamefont{Almheiri, Mahajan,
  and Maldacena}}]{Almheiri19}
\bibinfo{author}{\bibfnamefont{A.}~\bibnamefont{Almheiri}},
  \bibinfo{author}{\bibfnamefont{R.}~\bibnamefont{Mahajan}}, \bibnamefont{and}
  \bibinfo{author}{\bibfnamefont{J.}~\bibnamefont{Maldacena}},
  \emph{\bibinfo{title}{Islands outside the horizon}} (\bibinfo{year}{2019}),
  \eprint{1910.11077}.

\bibitem[{\citenamefont{Okuyama}(2019)}]{Okuyama19}
\bibinfo{author}{\bibfnamefont{K.}~\bibnamefont{Okuyama}},
  \bibinfo{journal}{arXiv preprint arXiv:1903.11776}  (\bibinfo{year}{2019}).

\bibitem[{\citenamefont{Susskind}(2016)}]{Susskind16}
\bibinfo{author}{\bibfnamefont{L.}~\bibnamefont{Susskind}},
  \bibinfo{journal}{Fortschritte der Physik} \textbf{\bibinfo{volume}{64}},
  \bibinfo{pages}{72} (\bibinfo{year}{2016}),
  \urlprefix\url{https://onlinelibrary.wiley.com/doi/abs/10.1002/prop.201500094}.

\bibitem[{\citenamefont{Cardy}(1989)}]{Cardy89}
\bibinfo{author}{\bibfnamefont{J.~L.} \bibnamefont{Cardy}},
  \bibinfo{journal}{Nuclear Physics B} \textbf{\bibinfo{volume}{324}},
  \bibinfo{pages}{581 } (\bibinfo{year}{1989}), ISSN \bibinfo{issn}{0550-3213},
  \urlprefix\url{http://www.sciencedirect.com/science/article/pii/055032138990521X}.

\bibitem[{\citenamefont{Cardy}(2014)}]{Cardy14}
\bibinfo{author}{\bibfnamefont{J.}~\bibnamefont{Cardy}},
  \bibinfo{journal}{Phys. Rev. Lett.} \textbf{\bibinfo{volume}{112}},
  \bibinfo{pages}{220401} (\bibinfo{year}{2014}),
  \urlprefix\url{https://link.aps.org/doi/10.1103/PhysRevLett.112.220401}.

\bibitem[{\citenamefont{Cardy}(2016)}]{Cardy16}
\bibinfo{author}{\bibfnamefont{J.}~\bibnamefont{Cardy}},
  \bibinfo{journal}{Journal of Physics A: Mathematical and Theoretical}
  \textbf{\bibinfo{volume}{49}}, \bibinfo{pages}{415401}
  (\bibinfo{year}{2016}),
  \urlprefix\url{https://doi.org/10.1088%2F1751-8113%2F49%2F41%2F415401}.


\bibitem[{\citenamefont{HaMa}(2013)}]{HaMa13}
\bibinfo{author}{\bibfnamefont{T.}~\bibnamefont{Hartman}}
\bibnamefont{and}
  \bibinfo{author}{\bibfnamefont{J.}~\bibnamefont{Maldacena}},
  \bibinfo{journal}{JHEP} \textbf{\bibinfo{volume}{1305}},
  \bibinfo{pages}{014} (\bibinfo{year}{2013}),
  \urlprefix\url{https://link.springer.com/article/10.1007%2FJHEP05%282013%29014
}.

\bibitem[{\citenamefont{Ishibashi}(1989)}]{Ishibashi89}
\bibinfo{author}{\bibfnamefont{N.}~\bibnamefont{Ishibashi}},
  \bibinfo{journal}{Modern Physics Letters A} \textbf{\bibinfo{volume}{04}},
  \bibinfo{pages}{251} (\bibinfo{year}{1989}),
  \urlprefix\url{https://doi.org/10.1142/S0217732389000320}.




\bibitem[{\citenamefont{Das and Datta}(2015)}]{DasDatta15}
\bibinfo{author}{\bibfnamefont{D.}~\bibnamefont{Das}} \bibnamefont{and}
  \bibinfo{author}{\bibfnamefont{S.}~\bibnamefont{Datta}},
  \bibinfo{journal}{Phys. Rev. Lett.} \textbf{\bibinfo{volume}{115}},
  \bibinfo{pages}{131602} (\bibinfo{year}{2015}),
  \urlprefix\url{https://link.aps.org/doi/10.1103/PhysRevLett.115.131602}.


\end{thebibliography}

\end{document}